\documentclass{aastex}
\usepackage{spr-astr-addons}
\usepackage{url}
\usepackage{amsmath}

\usepackage{longtable}

\RequirePackage{color}

\begin{document}

\title{Broad-band spectral evolution and temporal variability of IGR J17091$-$3624 during its 2016 outburst : {\it SWIFT} and {\it NuSTAR} results}
\slugcomment{Not to appear in Nonlearned J., 45.}
\shorttitle{Spectral evolution and variability of IGR J17091$-$3624}
\shortauthors{Radhika D. et al.}

\author{Radhika D.\altaffilmark{1}} \author{H. Sreehari\altaffilmark{2,3} } \author{Nandi A.\altaffilmark{2}} \author{Iyer N.\altaffilmark{4} } \author{Mandal S.\altaffilmark{5} }
\altaffiltext{1}{Department of Physics, Dayananda Sagar University, Bangalore, 560068, India}
\altaffiltext{2}{Space Astronomy Group, SSIF/ISITE Campus, ISRO Satellite Center, Outer Ring Road,
Marathahalli, Bangalore, 560037, India}
\altaffiltext{3}{Indian Institute of Science, Bangalore, 560012, India}
\altaffiltext{4}{Albanova University Centre, KTH PAP, Stockholm, 10691, Sweden}
\altaffiltext{5}{Indian Institute of Space Science and Technology, Thiruvananthapuram, Kerala, India}

\begin{abstract}

We report on the 2016 outburst of the transient Galactic Black Hole candidate IGR J17091$-$3624 based on the observation campaign carried out with \emph{SWIFT} and \emph{NuSTAR}. 
The outburst profile, as observed with \emph{SWIFT-XRT}, shows a typical `q'-shape in the Hardness Intensity Diagram (HID). 
Based on the spectral and temporal evolution of the different parameters, we are able to identify all the spectral states in the q-profile of HID and the Hardness-RMS diagram (HRD). 
Both \emph{XRT} and \emph{NuSTAR} observations show an evolution of low frequency Quasi periodic oscillations (QPOs) during the low hard and hard intermediate states of the outburst rising phase. 
We also find mHz QPOs along-with distinct coherent class variabilities (heartbeat oscillations) with different timescales, similar to the $\rho$-class (observed in GRS 1915$+$105). 
Phenomenological modelling of the broad-band \emph{XRT} and \emph{NuSTAR} spectra also reveals the evolution of high energy cut-off and presence of reflection from ionized material during the rising phase of the outburst. 
Further, we conduct the modelling of X-ray spectra of \emph{SWIFT} and \emph{NuSTAR} in 0.5 - 79 keV to understand the accretion flow dynamics based on two component flow model. 
From this modelling, we constrain the mass of the source to be in the range of 10.62 - 12.33 M$_{\sun}$ with 90\% confidence, which is consistent with earlier findings.

\end{abstract}


\section{Introduction}
Galactic Black Hole (GBH) X-ray binaries (XRBs) are mostly observed in low mass 
X-ray binary (LMXB) systems. Only a few of the BH XRBs (Cyg X-1, LMC X-1 and LMC X-3) are found in high mass X-ray binary 
(HMXB) systems \citep{MR06}. These X-ray binaries are observed to exist as either persistent 
or transient \citep{Chen97,2016ApJS..222...15T,2016A&A...587A..61C} in nature. The 
persistent sources usually show consistently high X-ray luminosity ($\sim10^{37}$ erg sec$^{-1}$;\citet{Kuz97}) for a long duration 
\citep{1996ARA&A..34..607T}, except for sources like GRS 1915$+$105 which has aperiodic variability. Transient/outbursting sources remain quiescent for a 
long time and exhibit a sudden increase in X-ray flux (from mCrabs to 
12 Crab; \cite{1989ApJ...337L..81T}). The transients remain active for tens of days to a 
few months or a few years before returning back to quiescent phase where the X-ray flux becomes 
non-detectable \citep{MR06}. 

A detailed understanding of the X-ray emission features (i.e., spectral and temporal 
characteristics) of the transient BH source during its outburst, is very essential to 
know about the accretion dynamics around the vicinity of BH XRB. 
Most of the BH transients usually exhibit thermal and non-thermal emission in their X-ray 
spectra. Thermal emission arises from the different radii of the Keplerian accretion disc which results in
 a multi-color blackbody spectrum at lower energies (i.e. soft spectrum) 
\citep{1973A&A....24..337S}. The non-thermal emission is due to Comptonization of disc 
photons by a static or dynamic hot corona existing in the innermost regions. 
This will result in a powerlaw spectral shape at higher energies (i.e. hard spectrum) usually with a cut-off 
\citep{1995xrbi.nasa..126T,ST95}. Sometimes due to the illumination of the disc by this non-thermal emission, a 
reflection component is also observed at higher energies \citep{RF93}. In general, the ratio of flux in a higher energy 
band (say 6 - 20 keV) to lower energy band (e.g. 2 - 6 keV) defines the hardness ratio 
\citep{Belloni2005,Nandi2012,RN2014}. During the outburst, the source intensity (i.e. X-ray flux) is 
observed to change with hardness ratio resulting in a `q'-shape plot which is well known as 
Hardness-Intensity Diagram (HID) 
(see \citealt{2001ApJS..132..377H,Belloni2005,Nandi2012,2016A&A...587A..61C} and references 
therein). Temporal analysis of the observations usually suggest that the GBH sources exhibit an evolution of the fractional rms variability during the outburst. Some times there are presence of low frequency QPOs which are classified into types A, B, C, C* based on their Q-factor, significance and amplitude \citep{Casella2004,MR06,2011BASI...39..409B}.

Depending upon the variation of the above mentioned spectral and temporal 
properties, the transient GBH sources occupy different spectral states in their HID. These states are classified as low hard (LHS), 
hard intermediate (HIMS), soft intermediate (SIMS) and high soft state (HSS). For details we refer to \citealt{2001ApJS..132..377H,FBG04,HB2005,Belloni2005,2006ARA&A..44...49R,Nandi2012,Motta2012} and references therein. Several works have been done based on the above spectral state classification, which has 
helped immensely to understand the spectral and temporal properties of BH sources and 
the evolution of their HID \citep{2001ApJS..132..377H,FBG04,FHB09,Belloni2005,MR06,Nandi2012,RN2014,RNVS16b}. In this paper, we refer to 
this general understanding of spectral state classification.

In addition to these characteristics which are generally observed in BH LMXBs, some sources show different types of 
variabilities/oscillations. These are usually referred to as coherent variabilities which may appear in the form of quasi-periodic flares or dips which occur for time 
period of seconds to minutes. The BH binaries GRS 1915$+$105 \citep{2001A&A...372..551B} and IGR J17091-3624 \citep{Alt11} exhibit 
these oscillations/variabilities. They are usually segregated into different classes because of the difference in X-ray flux, periodicity etc. 
The GBH transient source IGR J17091$-$3624 was discovered by 
\emph{International Gamma-ray Astrophysics Laboratory (INTEGRAL)} \citep{Kuul03} 
during 2003. Prior to this it appeared as a moderately bright transient during the 
period of 1994 to 2001 \citep{2003ATel..160....1I}. Thus the source has undergone multiple 
outbursts (2003, 2007 and 2011) till date. Detailed study of the spectral and 
temporal properties of the source suggests that it is similar to GRS 1915$+$105 \citep{Alt11}. 
Both sources exhibit coherent X-ray variability classes (heartbeat oscillations) at 
lower flux values, spectral state transitions and high frequency QPOs 
\citep{1999ApJ...527..321M,2001A&A...372..551B,Alt11,AB2012,Cap12,Zhang2014}. 

IGR J17091$-$3624 had 
undergone state transitions during its 2011 outburst, and variabilities/oscillations in 
timescales of 100 sec were observed in the light curves. These X-ray variability 
signatures were classified into $\nu$, $\rho$, $\alpha$, $\lambda$, $\beta$, $\mu$, $\gamma$ and $\chi$ \citep{Alt11,Zhang2014,2017arXiv170309572C} and observed to be similar with GRS 1915$+$105. During the time when the 
light curve displayed variabilities in IGR J17091$-$3624, the source had a softer spectra 
but exhibited high rms variability. The evolution of the spectral states 
and the oscillations observed in 2011 are not similar to the previous outbursts in 2003 and 
2007 where the source characteristics resembled with typical BH sources \citep{Cap12,Cap13}. 
Although there is no published literature which 
discusses about a complete HID of the source in 2011 outburst, the observations by \citealt{PahariATEL1,PahariATEL2} (ATEL 4282 and 4283) have shown a 
decline in source flux towards the quiescence. Since the \emph{XRT} observations had weak signal-to-noise ratio, \citealt{PahariATEL1} 
could not perform the detailed spectral analysis.

Recently \citealt{Xu17} have studied the rising phase of 2016 outburst of this source and looked into the spectral and 
temporal characteristics. They have discussed about reflection features and QPOs from the {\it NuSTAR} spectra for the rising phase of the outburst.

Even though IGR J17091$-$3624 is being considered as similar to GRS 1915$+$105, 
an estimate of its dynamical mass has not yet been obtained 
unlike GRS 1915$+$105. Likewise, the distance to the source IGR 17091$-$3624 and the disc 
inclination could not be determined due to lack of observational 
evidence of the nature of its binary companion. Previous attempts to estimate the mass of 
the source suggest the value to vary between 3 M$_{\sun}$ and 15 M$_{\sun}$ 
\citep{Alt11,RV2012, Rebusco12, AB2012, Pah2014}. A recent estimate points out a probable 
range for the mass as 8.7 M$_{\sun}$ to 15.6 M$_{\sun}$ 
\citep{2015ApJ...807..108I} based on spectral and temporal 
modelling, and 11.8 M$_{\sun}$ to 13.7 M$_{\sun}$ by 
modelling the broad-band energy spectra alone. The source is estimated to be at a 
distance of 10 kpc to 20 kpc by \cite{Alt11}. A better constraint of 
11 kpc to 17 kpc is given by \cite{rodriguez2011first} for a black hole of 
mass 10 M$_{\sun}$ using estimated luminosity at the hard to soft state transition. 
The inclination of IGR J17091$-$3624 has been proposed to be between 50$\degr$ to 70$\degr$ 
by \cite{king2012extreme} as disc-winds are present only in systems with high 
inclination angles. But it has to be noted that the inclination cannot exceed 70$\degr$ due to the absence of 
any signature of eclipses. Most of the mass estimates depend on the assumptions of
inclination and distance. This leads to the large spread in the range
of possible values. Thus it is difficult to know a precise value of mass
from these methods unless the inclination and distance are known accurately.
However, as stated in section \ref{ss:bbspec} the mass modelling method based on two 
component flow has little dependency on inclination or distance.

The source IGR J17091$-$3624 went into outburst during early 2016 and was detected 
by \emph{SWIFT-Burst Alert Telescope (BAT)} \citep{2016ATel.8742....1M}. The BAT 
light-curve shows a fast rise and exponential decay profile, extending from MJD 57445 
(27th Feb 2016) to 57615 (15th August 2016). The INTEGRAL observations 
\citep{2016ATel.8761....1G} indicated the source to be in its hard state 
during the rising phase of the outburst. Spectral transition to the intermediate 
state was observed during 22nd March i.e. 
MJD 57469 \citep{2016ATel.8858....1C} based on \emph{SWIFT} observations. 
During 13th April 2016 (MJD 57491), `heartbeat' oscillations have been detected with frequency of $\sim$ 0.027 Hz, 
using \emph{SWIFT-X-ray Telescope (XRT)} observations \citep{2016ATel.8948....1R}. The 
corresponding X-ray spectrum has been understood to consist of emission due to 
both Keplerian disc (thermal) and Comptonized emission from the corona (non-thermal). 
Optical observation has found the source magnitude to be brighter by 1.5 in all the bands 
\citep{2016ATel.8795....1G} in comparison to the magnitude value in 2011 outburst. 
There has been no detection of any jet ejection 
in the radio band from this source during the 2016 outburst \citep{2016ATel.8821....1E}.

In this paper, we consider \emph{SWIFT-XRT} and \emph{Nuclear Spectroscopic Telescope Array 
(NuSTAR)} observations for the 2016 outburst of the source IGR J17091$-$3624. We explore 
the spectral and temporal characteristics of the source, so as to look for the spectral state 
transitions during this outburst. The evolution of HID of the source is studied based on 
the phenomenological models to understand the contribution of the soft and hard 
components separately. We search for the evidence of coherent oscillations/variabilities in the 
light-curve, and how they evolve as the outburst progresses. Then we attempt to see 
whether these variabilities have any correlation with the different spectral states. 
The characteristics of PDS are also being looked into so as to understand the evolution 
of low frequency QPOs during the rising phase of the outburst. Finally, based on the 
two component accretion flow paradigm, we model the energy spectra of the four 
quasi-simultaneous broad-band (0.5 - 79 keV) observations using \emph{SWIFT} and \emph{NuSTAR}. 
Rest of the {\it XRT} data (51 in number, spanning over 172 days) and the 
two \emph{NuSTAR} data which are not taken simultaneously with SWIFT are also modelled separately in the 
same way. The procedure for modelling is based on \citealt{2015ApJ...807..108I}. 
From this, we understand the variations of the model parameters during the different 
spectral states. We generate the HID from phenomenological fits and perform a 
comparative study with respect to the results obtained from two component model fits. We also 
constrain the mass of the source from the two component model fitting of energy spectrum 
from different spectral states. We also construct the probability distribution function of the source mass, for having a better constrain on the mass.

A summary of the procedures followed for data reduction has been given in 
section \ref{obs}. The methodology considered for analysis of data from \emph{XRT and NuSTAR} are discussed in section \ref{anal}. 
The results obtained from the spectral and temporal analysis using 
phenomenological and two component flow model are presented in section \ref{res}. These results
have been discussed in section \ref{dis}. 

\section{Observations and data reduction}
\label{obs}
We have analysed the public archival data of the \emph{SWIFT} satellite, available through 
the HEASARC database. Data are obtained for 51 observations beginning 
from the first day of outburst i.e. MJD 57445 (27th Feb 2016) and up-to 
MJD 57617 (17th August 2016) when the source is in its decay phase. Six Target Of 
Opportunity (TOO) observations of \emph{NuSTAR} among which four are quasi-simultaneous 
with \emph{SWIFT} are also considered. An observation log has been tabulated as part of the 
appendix (see Table \ref{tab:obs}). In this paper, we refer \textit{to MJD 57445.0 as day 0}, and 
all the other observations follow accordingly. The standard ftools provided 
by \texttt{HEASOFT v 6.20} are used for the purpose of data reduction and analysis. 

\subsection{\emph{XRT} data reduction}
\emph{SWIFT-XRT} \citep{2005SSRv..120..165B} 
has observed the source IGR J17091$-$3624 using window-timing mode. This
has data covering the energy range of 0.2 - 10 keV. 
The cleaned {\it XRT} event products are obtained through the \texttt{xrtpipeline} and events are 
selected \footnote{http://www.swift.ac.uk/analysis/xrt/xrtpipeline.php} corresponding 
to grades of \emph{0-2} using \texttt{XSELECT v 2.4}. 

\begin{figure}
\includegraphics[height=6cm,width=8.2cm]{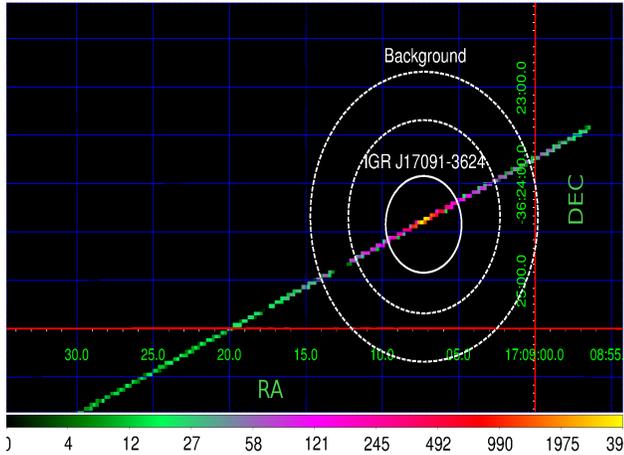}
\caption{Image of the source IGR J17091$-$3624 taken with \textit{SWIFT-XRT} and displayed in 
logarithmic scale. The dotted annular region has inner radius 60$\arcsec$ and outer radius 
of 90$\arcsec$ while the circular source region has a radius of 30$\arcsec$. Both regions 
are centred at the RA and Dec of the source.}
\label{fig:XRTimage}
\end{figure}

As per the XRT threads \footnote{http://www.swift.ac.uk/analysis/xrt/pileup.php}, if the Window-timing mode data has more than 100 counts/sec then pile-up may occur in the imaging. For the observations of this source the count rate is less than 30. Hence the image is completely devoid of pile-up effect. We choose 
a circle of radius 30$\arcsec$ for the source region and an annular region is taken far away 
from the source for the background, as shown in Figure \ref{fig:XRTimage}. The x-axis is Right Ascension (RA) and y-axis is Declination (DEC). We have also provided a color-bar at the bottom of the figure indicating the intensity. We apply a scaling 
factor by editing the \texttt{BACKSCAL} keyword 
\footnote{http://www.swift.ac.uk/analysis/xrt/backscal.php} for both source and 
background regions (see \citealt{RNVS16b} for details). 

Uncertainty in position of the source has been taken care of, by applying the 
position dependent \emph{rmfs} for the grade \emph{0-2}. The ARF files are obtained by 
making use of the exposure map with \texttt{xrtmkarf}. We re-bin the 
source spectral data to contain a minimum of 25 counts per bin with the ftool \texttt{grppha}. 

The extraction of {\it XRT} timing data is performed by following the procedure mentioned 
in the {\it XRT} analysis guide \footnote{http://www.swift.ac.uk/analysis/xrt/timing.php}. We 
consider the event data of {\it XRT} and select the source and background regions. \footnote {It has to be noted that for observations where the XRT image showed double streaks, we excluded the events corresponding to the streak having shorter good time interval.} For the 
respective regions, we generate source and background light curves. A time bin resolution 
of 0.018 sec is chosen as multiple of the minimum {\it XRT} time resolution of 1.8 msec, while 
obtaining the light curves. With the help of \emph{lcmath}, we subtract the background 
light curve from the source. Thus we obtain the background subtracted light 
curve, which is used for further analysis. Detailed procedures maybe referred to in \citealt{RNVS16b}.

\subsection{\emph{NuSTAR} data reduction}
The \emph{NuSTAR} mission \citep{harrison2013nuclear} consists of two independent grazing 
incidence telescopes operating in the high energy X-rays (3 - 79 keV). 
It has two Focal Plane Modules (FPM) referred to as FPMA and FPMB. \emph{NuSTAR} has six 
TOO observations of IGR J17091$-$3624 which are carried out over different phases of 
the outburst. Since four \emph{NuSTAR} observations are quasi-simultaneous with \emph{SWIFT}, 
we could obtain broad-band spectra for the energy range of 0.5 - 79 keV. We have 
got statistically sufficient counts for these broad-band spectra, in all the 
states of the source as illustrated in Table \ref{tab:obs} of appendix. 

\begin{figure}
\includegraphics[height=6cm,width=8.2cm]{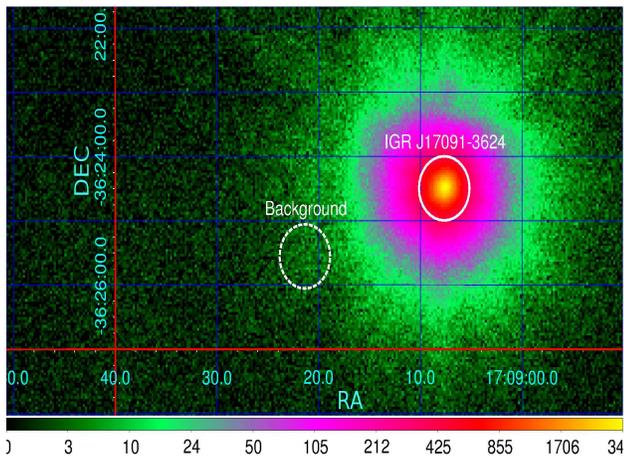}
\caption{Image shows IGR J17091$-$3624 captured with \emph{NuSTAR} displayed in logarithmic 
scale. The source region is a circle of radius 30$\arcsec$ centred at the RA and Dec of the 
source while the background is a dashed circle of 30$\arcsec$ radius chosen slightly away 
from the source as is suggested in the \emph{NuSTAR} quickstart guide.} 
\label{fig:Nustarimage}
\end{figure}

Data from both FPM detectors of the \emph{NuSTAR} observatory is used for obtaining spectra 
between 3 - 79 keV. We follow the procedures mentioned in the \emph{NuSTAR} 
guide \footnote{https://heasarc.gsfc.nasa.gov/docs/nustar/analysis/nustar\_swguide.pdf} 
and extract the level 2 data using the ftools command \texttt{nupipeline}. The source 
spectrum and light curve are extracted using a circular region of 30$\arcsec$ centred at 
the source RA and DEC with the $ds9$ tool. For background spectrum, we use another source
free circular region of 30$\arcsec$ (see Figure \ref{fig:Nustarimage}) taken from the same 
detector on which the source is seen. The ftool command  \texttt{nuproducts} is 
used to extract the spectrum, light curve, response and arf files. The spectrum so obtained 
is re-binned to contain a minimum of 30 counts per bin using $grppha$ in order to use $\chi^2$ statistics while 
fitting the data. 

\section{Methodologies for `spectro-temporal' analysis and phenomenological modelling}
\label{anal}
Using the spectral package of \texttt{XSpec v 12.9} \citep{Arn96}, we do spectral 
analysis of both \emph{XRT} and \emph{NuSTAR} data. We consider the energy ranges of 
0.5 - 10 keV (\emph{XRT}) and 3 - 79 keV (\emph{NuSTAR}) respectively for the entire observation, as the optimum energy 
range with statistically significant photon counts. In this section we summarize the methodologies used to analyze spectral and temporal data, from both \emph{XRT and NuSTAR}.

\subsection{Analysis and modelling of \emph{XRT} observations}

We perform the spectral modelling using the \emph{diskbb} \citep{1984PASJ...36..741M,Maki86} 
and \emph{powerlaw} models. These models take into account the low-energy thermal 
emission from the Keplerian accretion disc and 
the Comptonization of the disc photons by the corona respectively. The \emph{phabs} 
model \citep{2000ApJ...542..914W} is used to consider the interstellar absorption. 
From the spectral fits to all the data sets, the hydrogen column density (n$_H$) is 
found to be of the order of 1.1$\times$10$^{22}$ cm$^{-2}$. This value is similar to the 
estimates of \citealt{2011ATel.3144....1K,Cap12}. 

With the \emph{cflux} model, we obtain the unabsorbed X-ray flux from the source in the
{\it XRT} energy band of 0.5 - 10 keV. The {\it XRT} hardness ratio is estimated by calculating the 
ratio of fluxes in 4 - 10 keV and 0.5 - 4 keV. It is used to plot the HID along-with 
the total flux in the energy range of 0.5 - 10 keV. 
Using the \textit{err} command of \texttt{XSpec}, we find the error limits for the flux values and 
different spectral parameters at 90 percent confidence interval. Here, 
the parameter for which error limits are to be obtained varies within a specific already assigned limit. This continues until the value of fit statistic 
becomes greater than the fit statistic obtained in the preceding step by a specific amount \citep{Arn96}.  

The package \texttt{XRONOS v 5.22} is used for generating temporal {\it XRT} data with time 
resolution of 0.018 sec and bin size of 8192. This corresponds to a frequency range 
of 0.007 - 27.8 Hz for the PDS. Using \texttt{powspec v 1.0}, we generate PDS for the 
background subtracted light-curves so as to search for the presence of very low frequency 
QPOs. The PDS are normalized so that the integral gives the squared fractional rms 
variability. We also subtract the Poisson noise which is found to be a flat spectra 
with a power value of $\sim$2. Thus the resulting PDS is expressed as rms power (in units of rms$^2$/Hz) variation with 
frequency\footnote{https://heasarc.gsfc.nasa.gov/ftools/fhelp/powspec.txt}. 

The fractional rms variability is estimated for the frequency range of 0.007 - 1 Hz, since the noise is 
dominant above 1 Hz for most of the observations. As an example let us consider the initial observation of the source 
on day 0. The rms variability obtained for the frequency range of 0.007 - 1 Hz gives a value 39.3\%. For 0.007 - 2 Hz 
the variability increases to 48\%, while for 0.007 - 10 Hz it is $>$ 100\%. Thus we find that above 1 Hz the rms variability is 
significantly dominated by band-limited noise. The rms value is calculated using the rectangle rule integration method, following the procedure given in RXTE cookbook \footnote{https://heasarc.gsfc.nasa.gov/docs/xte/recipes/pca\_fourier.html}. 

As the count rates are low ($<$ 30 counts/sec), we divide the entire light curves into intervals of 2048 bins to 
obtain individual power spectrum. Then we co-add all these power spectra and average them to 
get statistically significant detection of QPOs. The different components of 
the PDS are fitted with multiple Lorentzians or else with a constant factor for noise 
dominated frequency range. The centroid 
of this Lorentzian ($\nu$) gives QPO frequency. Coherence of a QPO is defined by the Q-factor 
which is the ratio of centroid frequency ($\nu$) to width ($\lambda$). The QPO significance 
is calculated as the ratio of Lorentzian normalization to its negative error. QPO amplitude 
refers to the integrated rms power in the Lorentzian which is estimated using the rectangle 
rule integration method as mentioned earlier. 
For detailed procedure see \citealt{BH1990,Casella2004,MR06}.  \\

\subsection{Analysis and modelling of \emph{NuSTAR} observations}

The source IGR J17091$-$3624 during its 2016 outburst has been observed by \emph{NuSTAR} on 
MJDs 57454, 57459, 57461, 57476, 57508 and 57534. We analyze all the six \emph{NuSTAR} 
observations using both FPMA and FPMB detectors. The spectral fit parameters for 
data with both detectors is found to be similar. 
Hence, we present in this paper the analysis based on FPMA data only. It has to be also noted 
that the entire \emph{NuSTAR} analysis has been done without inclusion of the fluorescent Fe 
line emission.

Four among the six TOO observations aid in the broad-band study of state evolution 
of the source. The broad-band spectrum in the energy range of 0.5 - 79 keV from 
\emph{SWIFT-XRT} and \emph{NuSTAR} is well fitted by the phenomenological models 
consisting of \textit{diskbb}, \textit{ireflect} and \emph{cutoffpl}. The \emph{NuSTAR} 
has excellent timing capability with a temporal resolution of 0.1 ms 
\citep{harrison2013nuclear}. Since we intend to search for the presence of low frequency QPOs, 
background subtracted light curves of resolution 0.5 sec are generated. The light curves are then divided into
 8192 intervals, to result in a frequency range of 0.0002 - 1 Hz in the PDS. Here, 
also we model the different components of the PDS and hence study the properties 
of QPOs. Background subtracted light curves of same bin time are also 
generated for different energy bands (3 - 15 keV, 15 - 79 keV and 3 - 79 keV). Light curves 
are used to produce PDS for various energy bands and fitted with Lorentzians to extract 
different features, as mentioned above. 

All these procedures discussed for analysis of the data are based on phenomenological 
models. In the following section, we will discuss in detail the results from our temporal 
and spectral modelling of the \emph{XRT} and \emph{NuSTAR} observational data 
for the entire outburst.

\section{Results}
\label{res}
We present here the results of detailed spectral and temporal analysis of the source 
IGR J17091$-$3624 during its 2016 outburst. 
In subsection \ref{sptmpe} we give the details of results obtained using phenomenological fits to the X-ray spectra, and from temporal analysis. The evolution of the different HID regions are discussed here. We also give a description of evidence obtained for coherent variabilities and evolution of QPOs during some of the HID regions in \ref{coh-qpo}. 
Further, we give an account of the spectral modelling performed with two component model and the variations of relevant parameters.

\subsection{Evolution of spectral and temporal features in the outburst profile}
\label{sptmpe}

\begin{figure}
\begin{center}
\includegraphics[width=10cm,height=14cm]{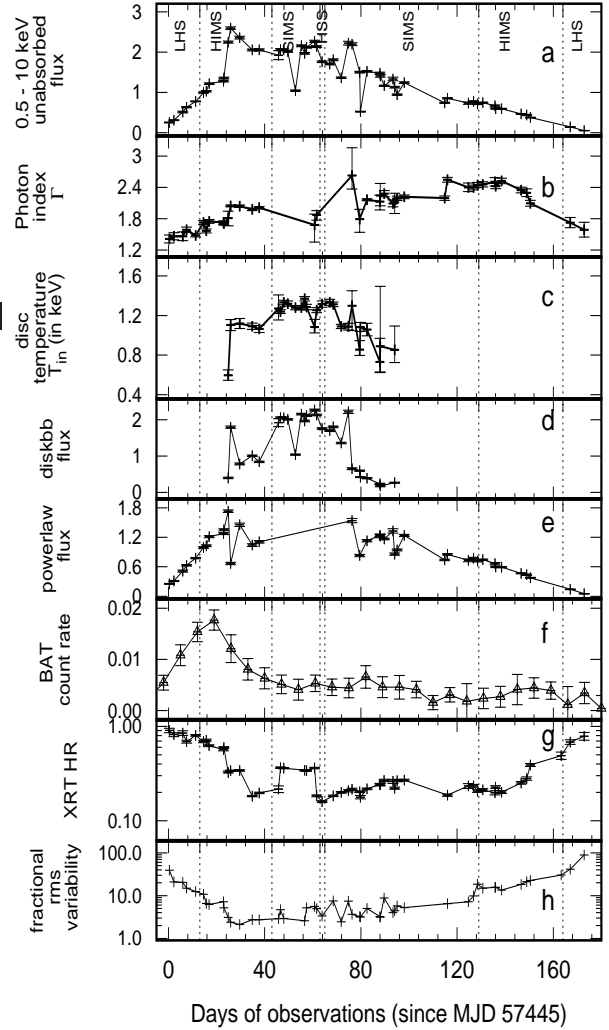}
\caption{Evolution of total X-ray flux in 0.5 - 10 keV, spectral parameters of photon index and disk temperature, contribution of disc flux and 
powerlaw flux in 0.5 - 10 keV, BAT count rate in 15 - 50 keV, hardness ratio and fractional rms variability 
in \% computed in 0.007 - 1 Hz frequency range are presented in different panels. 
All the flux values are given in units of $10^{-9}$ erg cm$^{-2}$ sec$^{-1}$. The grids 
(vertical lines) are used to differentiate between the different spectral states. See text 
for details.}
\label{fig:fig1}
\end{center}
\end{figure}

\begin{figure*}
	\begin{minipage}{8cm}
		\includegraphics[width=8cm]{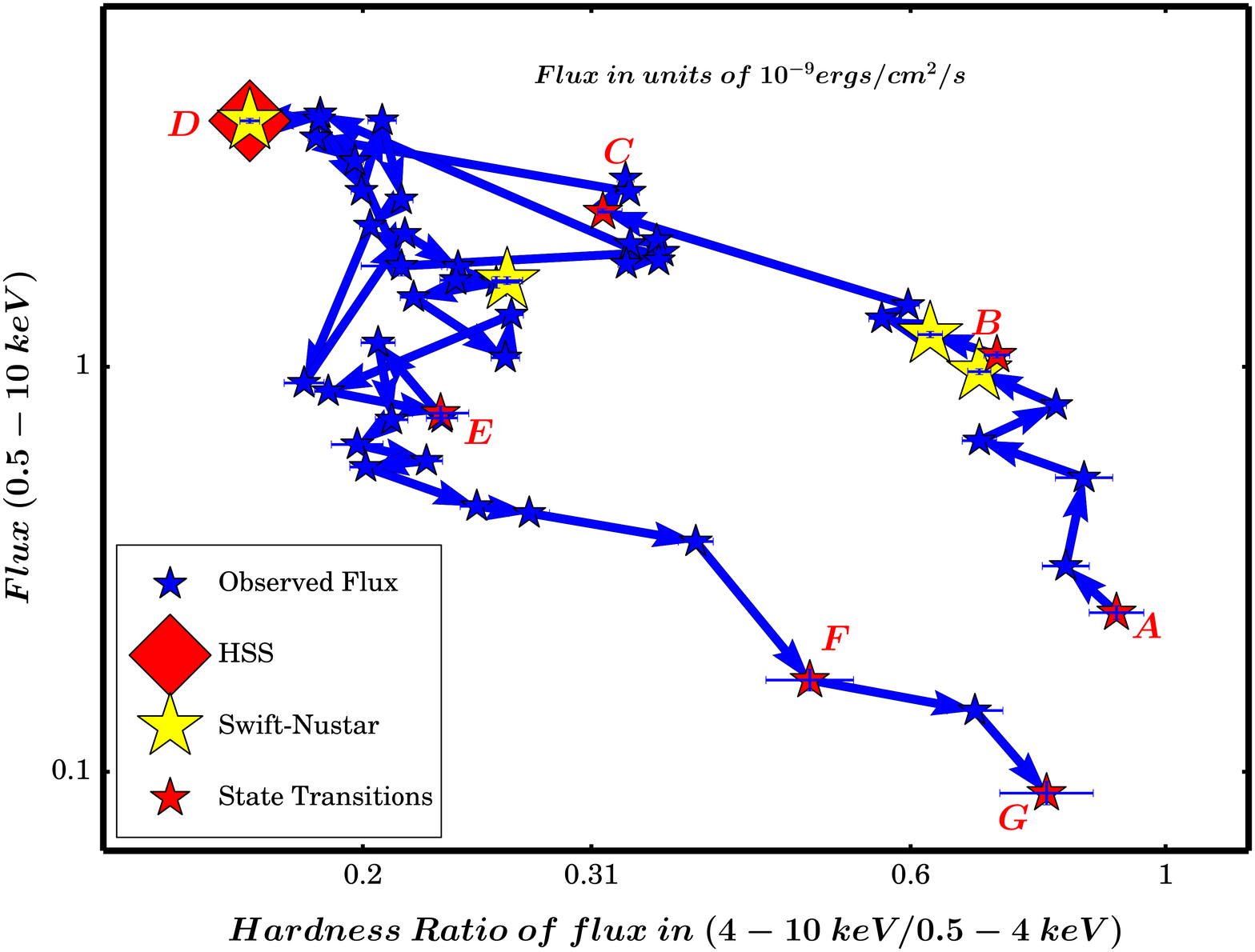}
	\end{minipage}
	\begin{minipage}{8cm}
		\includegraphics[width=8cm]{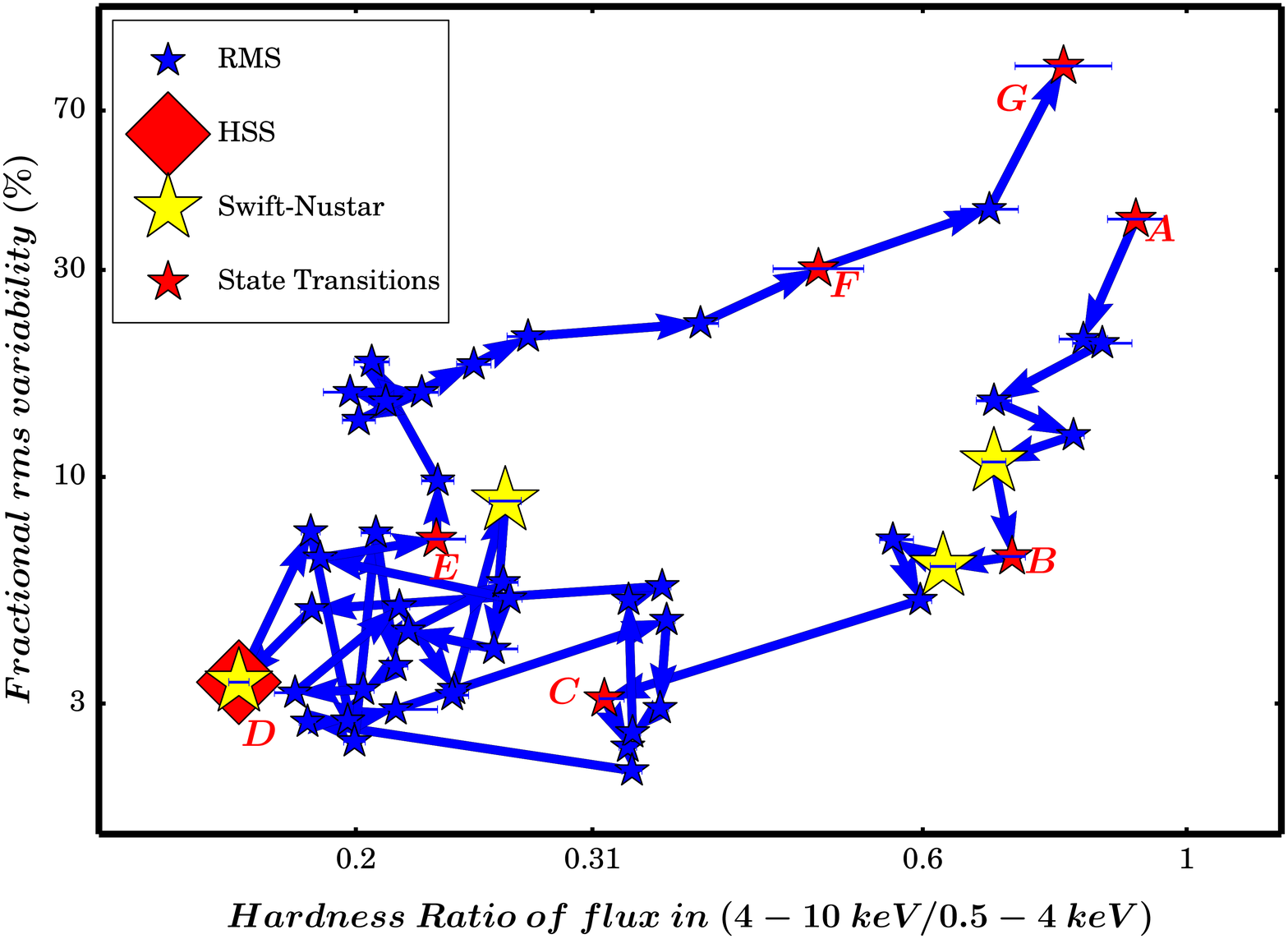}
	\end{minipage}
\caption{Left panel : Hardness-Intensity diagram (HID) showing evolution of the hardness ratio (flux in 4 - 10 keV / 0.5 - 4 keV) with total 
unabsorbed flux in 0.5 - 10 keV given in units of $10^{-9}$ erg cm$^{-2}$ sec$^{-1}$ is 
shown. Right panel : Hardness-rms diagram (HRD) depicting evolution of hardness ratio w.r.t fractional rms variability (in \%) estimated in a frequency range of 0.007 - 1 Hz. The different regions in the HID \& HRD from A to G denote the spectral states of hard, 
hard intermediate, soft intermediate and soft state during the rising and decaying phase 
of the outburst respectively. The four quasi-simultaneous broad-band observations with \emph{XRT+NuSTAR} are 
marked in yellow stars.}
\label{fig:fig3}
\end{figure*}

In Figure \ref{fig:fig1}, we present the 
evolution of observed flux. The unabsorbed value of flux in 0.5 - 10 keV, along-with the contribution of disc and powerlaw
 fluxes are shown in panels a, d and e respectively. The variation of spectral parameters i.e. photon index $\Gamma$ and disc temperature T$_{in}$ are represented in panels b and c. The evolution of BAT count rate obtained from BAT light curve of the 
source\footnote{https://swift.gsfc.nasa.gov/results/transients/} is shown in panel f. The XRT hardness ratio and fractional rms 
variability obtained from power spectra for all observations are represented in panels g and h respectively. 
The evolution of total flux as a function of hardness ratio for {\it XRT} observations is 
shown in the hardness intensity diagram (HID) in left panel of Figure \ref{fig:fig3}. On the right panel of Figure \ref{fig:fig3} we show the variation of the fractional rms variability w.r.t. the hardness ratio in a hardness rms diagram (HRD). Phenomenological fitting performed for broad-band spectra of quasi-simultaneous \emph{XRT and NuSTAR} 
observations are also discussed in this section and shown in Figure \ref{fig:phenoss}.

In this section we also present the results on unique temporal 
properties of the source. In Figure \ref{fig:LCPDS}, we show the variation of {\it XRT} 
light-curves and power spectra for various phases of the outburst. The \emph{NuSTAR} power spectra are presented in 
Figure \ref{fig:Nustar-PDS}. Different 
types of variabilities that are observed in {\it XRT} light-curves are shown in 
Figure \ref{fig:Variability}.

Below, we summarize the evolution of spectral and temporal parameters during the 
different regions of the HID and HRD (marked from A to G in Figure \ref{fig:fig3}) as the outburst
progresses. As mentioned in section 1, we refer to the general understanding of
spectral state classification \citep{Belloni2005,MR06,Nandi2012} while studying
the HID. In Table \ref{tab:parameters} we have given the values of different spectral and temporal parameters 
obtained from {\it XRT} observations, while Table \ref{tab:qpo} presents the results from temporal analysis of specific {\it XRT} 
and {\it NuSTAR} observations. Tables \ref{tab:phenofit} and \ref{tab:resfit} show results from spectral analysis of broad-band 
observations from the different HID regions. We bring into notice once again that MJD 57445 is considered as day 0 throughout the entire manuscript. 

All regions discussed in the following sections correspond to regions of Figure \ref{fig:fig3}.  

\begin{table*}
\caption{Variation of spectral and temporal parameters obtained from {\it XRT} observations. It has to be noted that the X-ray flux values are quoted here in units of $10^{-9}$ erg cm$^{-2}$ sec$^{-1}$, and the fractional rms variability has been estimated in the frequency range of 0.007 - 1 Hz}.
\label{tab:parameters}
\begin{tabular}{ccccccc}
\hline
Spectral state & MJD & 0.5 - 10 keV flux & Hardness Ratio & Photon index, $\Gamma$ & T$_{in}$ & rms(\%)\\
\hline
LHS-rise 
&57445.35	&2.524$^{+0.09}_{-0.09}$	&0.906$^{+  0.049 }_{-0.049	}$	&1.411$^{+0.076}_{-0.096}$	&-	&39.314\\
&57447.27	&3.105$^{+0.104}_{-0.103}$	&0.818$^{+  0.038 }_{-0.038	}$	&1.463$^{+0.068}_{-0.068}$	&-      &20.793\\
&57450.94	&5.122$^{+0.206}_{-0.206}$	&0.849$^{+  0.048 }_{-0.048	}$	&1.462$^{+0.083}_{-0.083}$	&-      &20.343\\
&57452.53	&6.306$^{+0.153}_{-0.153}$	&0.688$^{+  0.023 }_{-0.023	}$	&1.587$^{+0.050}_{-0.050}$	&-      &14.985\\
&57456.26	&7.755$^{+0.113}_{-0.113}$	&0.803$^{+  0.016 }_{-0.016	}$	&1.479$^{+0.030}_{-0.030}$	&-      &12.475\\
\hline
HIMS-rise 
&57459.57	&9.940$^{+0.167}_{-0.166}$	&0.688$^{+  0.015}_{- 0.015}$	&1.727$^{+0.037}_{-0.036}$	&-              &10.828\\
&57460.71	&1.028$^{+0.018}_{-0.018}$	&0.713$^{+  0.018 }_{-0.018}$	&1.566$^{+0.037}_{-0.038}$	&-              &6.553\\
&57461.90	&1.214$^{+0.019}_{-0.019}$	&0.623$^{+  0.015}_{- 0.015}$	&1.745$^{+0.036}_{-0.036}$	&-              &6.218\\
&57467.88	&1.276$^{+0.020}_{-0.019}$	&0.566$^{+  0.014}_{- 0.022}$	&1.727$^{+0.032}_{-0.032}$	&-              &7.168\\
&57468.02	&1.355$^{+0.021}_{-0.021}$	&0.596$^{+  0.012}_{- 0.013}$	&1.706$^{+0.032}_{-0.032}$	&-              &5.201\\
&57469.81	&2.245$^{+0.026}_{-0.026}$	&0.323$^{+  0.003}_{- 0.012}$	&1.813$^{+0.151}_{-0.131}$	&0.596$^{+0.049}_{-0.056}$                     &3.081\\
&57470.81	&2.595$^{+0.031}_{-0.030}$	&0.338$^{+  0.005}_{- 0.005}$	&2.044$^{+0.016}_{-0.018}$	&1.104$^{+0.056}_{-0.056}$                        &2.387\\
&57474.60	&2.357$^{+0.032}_{-0.032}$	&0.341$^{+  0.006}_{- 0.006}$	&2.039$^{+0.022}_{-0.023}$	&1.119$^{+0.051}_{-0.051}$                        &2.104\\
&57479.84	&2.054$^{+0.023}_{-0.0234}$	&0.182$^{+  0.002}_{- 0.002}$	&1.978$^{+0.023}_{-0.024}$	&1.092$^{+0.032}_{-0.031}$                        &2.724\\
\hline
SIMS-rise 
&57482.69	&2.054$^{+0.024}_{-0.024}$	&0.196$^{+  0.003}_{- 0.003	}$	&2.009$^{+0.022}_{-0.023}$	&1.064$^{+0.036}_{-0.035}$                        &2.741\\
&57490.74	&1.923$^{+0.109}_{-0.109}$	&0.216$^{+  0.017 }_{-0.017	}$	&-				&1.269$^{+0.115}_{-0.137}$                        &2.91\\
&57491.54	&2.061$^{+0.024}_{-0.023}$	&0.365$^{+  0.005 }_{-0.005	}$	&-				&1.239$^{+0.019}_{-0.019}$                        &4.678\\
&57492.95	&2.056$^{+0.023}_{-0.023}$	&0.360$^{+  0.005}_{- 0.005	}$	&-				&1.332$^{+0.021}_{-0.022}$                       &2.929\\
&57501.64	&1.967$^{+0.021}_{-0.021}$	&0.341$^{+  0.005 }_{-0.005}$	&-				&1.369$^{+0.021}_{-0.021}$                       &2.576\\
&57502.38	&2.118$^{+0.023}_{-0.023}$	&0.339$^{+  0.005 }_{-0.005}$	&-				&1.285$^{+0.026}_{-0.027}$                        &5.207\\
&57505.77	&2.270$^{+0.025}_{-0.025}$	&0.362$^{+  0.006}_{- 0.006}$	&1.676$^{+0.324}_{-0.210}$	&1.082$^{+0.057}_{-0.078}$                       &5.596\\
&57506.58	&2.137$^{+0.023}_{-0.023}$	&0.183$^{+  0.004}_{- 0.004}$	&1.869$^{+0.099}_{-0.084}$	&1.254$^{+0.025}_{-0.026}$                       &4.952\\
\hline
HSS-rise 
&57508.89	&1.759$^{+0.021}_{-0.021}$	&0.159$^{+  0.003}_{- 0.003}$	&-				&1.314$^{+0.030}_{-0.031}$                       &3.363\\
\hline
SIMS-decay 
&57512.02	&1.701$^{+0.019}_{-0.019}$	&		&-				&1.333$^{+0.029}_{-0.031}$                       &4.704\\
&57513.47	&1.808$^{+0.023}_{-0.023}$	&0.183$^{+  0.003}_{-  0.003	}$	&-				&1.312$^{+0.023}_{-0.024}$                       &7.485\\
&57516.81	&1.362$^{+0.019}_{-0.019}$	&0.199$^{+  0.004}_{-  0.004	}$		&-				&1.090$^{+0.020}_{-0.021}$                       &2.454\\
&57519.79	&2.220$^{+0.044}_{-0.044}$	&0.207$^{+  0.005}_{-  0.005	}$	&-				&1.087$^{+0.032}_{-0.034}$                       &7.437\\
&57521.31	&2.193$^{+0.036}_{-0.036}$	&0.216$^{+  0.005}_{-  0.005	}$	&2.628$^{+0.263}_{-0.530}$	&1.299$^{+0.252}_{-0.149}$                        &3.663\\
&57524.50	&1.502$^{+0.022}_{-0.022}$	&0.202$^{+  0.004}_{-  0.004	}$	&1.791$^{+0.254}_{-0.186}$	&0.855$^{+0.059}_{-0.091}$                     &3.229\\
&57524.77	&5.216$^{+0.125}_{-0.125}$	&0.177$^{+  0.007}_{-  0.007	}$	&-				&1.082$^{+0.044}_{-0.047}$                       &3.18\\
&57527.56	&1.521$^{+0.021}_{-0.021}$	&0.217$^{+  0.004}_{-  0.004	}$	&2.163$^{+0.020}_{-0.021}$	&1.057$^{+0.062}_{-0.063}$                      &5.027\\
&57532.88	&1.477$^{+0.022}_{-0.022}$	&0.242$^{+  0.005}_{-  0.005	}$	&2.125$^{+0.159}_{-0.126}$	&0.730$^{+0.106}_{-0.238}$                       &3.241\\
&57533.03	&1.407$^{+0.031}_{-0.031}$	&0.241$^{+  0.007}_{-  0.007	}$	&2.244$^{+0.197}_{-0.231}$	&0.889$^{+0.259}_{-0.604}$                    &3.139\\
&57534.67	&1.164$^{+0.017}_{-0.017}$	&0.267$^{+  0.008}_{-  0.008	}$	&2.279$^{+0.059}_{-0.061}$	&-                                          		&8.80\\
&57538.35  	&1.330$^{+0.043}_{-0.043}$	&0.261$^{+  0.012}_{-  0.012	}$	&2.092$^{+0.064}_{-0.065}$	&-                                          		&4.012\\
&57539.07  	&1.127$^{+0.021}_{-0.021}$	&0.221$^{+  0.005}_{-  0.005	}$	&2.109$^{+0.212}_{-0.175}$	&0.854$^{+0.129}_{-0.239}$                    &4.417\\
&57540.07	&9.424$^{+0.190}_{-0.190}$	&0.266$^{+  0.007}_{-  0.007	}$	&2.188$^{+0.041}_{-0.041}$	&-                                          		&5.714\\
&57543.06	&1.240$^{+0.020}_{-0.020}$	&0.269$^{+  0.006}_{-  0.006	}$	&2.218$^{+0.035}_{-0.035}$	&-                                          		&5.241\\
\hline
HIMS-decay 
&57561.00	&8.523$^{+0.166}_{-0.166}$	&0.186$^{+  0.005}_{- 0.005}$	&2.545$^{+0.045}_{-0.046}$	&-                                          		&6.510\\
&57569.71	&7.321$^{+0.279}_{-0.279}$	&0.233$^{+  0.013}_{- 0.013}$	&2.399$^{+0.081}_{-0.082}$	&-                                          		&7.187\\
&57571.83	&7.768$^{+0.163}_{-0.163}$	&0.234$^{+  0.007}_{- 0.007}$	&2.397$^{+0.045}_{-0.045}$	&-                                          		&9.799\\
&57573.48	&7.151$^{+0.168}_{-0.167}$	&0.206$^{+  0.007}_{- 0.007}$	&2.446$^{+0.049}_{-0.049}$	&-                                          		&18.421\\
&57575.68	&7.442$^{+0.166}_{-0.166}$	&0.211$^{+  0.006}_{- 0.006}$	&2.456$^{+0.050}_{-0.051}$	&-                                          		&14.922\\
&57580.73	&6.644$^{+0.226}_{-0.227}$	&0.197$^{+  0.010}_{- 0.010}$	&2.507$^{+0.079}_{-0.081}$	&-                                          		&15.684\\
&57581.01	&5.962$^{+0.130}_{-0.130}$	&0.227$^{+  0.007}_{- 0.007}$	&2.424$^{+0.050}_{-0.051}$	&-                                          		&15.671\\
&57583.53	&5.934$^{+0.124}_{-0.124}$	&0.201$^{+  0.006}_{- 0.006}$	&2.524$^{+0.051}_{-0.052}$	&-                                          		&13.525\\
&57591.64	&4.667$^{+0.103}_{-0.103}$	&0.251$^{+  0.008}_{- 0.008}$	&2.348$^{+0.055}_{-0.055}$	&-                                          		&18.177\\
&57593.95	&4.381$^{+0.137}_{-0.137}$	&0.279$^{+  0.011}_{- 0.011}$	&2.295$^{+0.072}_{-0.073}$	&-                                          		&21.079\\
&57595.48	&3.752$^{+0.089}_{-0.089}$	&0.389$^{+  0.013}_{- 0.013}$	&2.095$^{+0.055}_{-0.055}$	&-                                          		&22.603\\
\hline
LHS-decay 
&57612.03	&1.413$^{+0.061}_{-0.061}$	&0.682$^{+  0.038}_{- 0.038}$	&1.723$^{+0.098}_{-0.098}$	&-                                          		&41.43\\
&57617.82	&5.183$^{+0.313}_{-0.313}$	&0.787$^{+  0.073}_{- 0.073}$	&1.587$^{+0.143}_{-0.143}$	&-                                          		&88.65\\

\hline
\end{tabular}
\end{table*}

\subsubsection{Region AB}

Observations during days 0.35 to 13.8 are considered part of the region AB. These observations belong to the rising phase of 
the outburst.

The \emph{XRT} spectra are well modelled with \emph{phabs*(powerlaw)} and do not 
require the \emph{diskbb}. As an example from the spectral fits on 
day 13.8, we obtain $\chi^{2}/dof$ as 360.01/340. An inclusion 
of \emph{diskbb} results in $\chi^{2}/dof$ of 358.99/338 but with incorrect values for the 
different \emph{diskbb} parameters. An ftest to this observation provides F-statistic of 0.48 and probability of 0.62. 
These prove that the disc component is not necessary for the fits.  

During these initial days, we observe the {\it XRT} flux and spectral photon index to increase as shown in 
Figure \ref{fig:fig1} and Table \ref{tab:parameters}. 
It is also evident that the {\it XRT} hardness ratio decreases from 0.906$\pm{0.049}$ to 
0.803$\pm{0.016}$ (see Table \ref{tab:parameters}, Figure \ref{fig:fig3} and panel g of Figure \ref{fig:fig1}). 
The BAT count rate in 15 - 50 keV corresponding to the hard 
flux contribution is also observed to increase from 0.007 to 0.017 (panel f of 
Figure \ref{fig:fig1}). The temporal analysis suggest that the fractional rms variability decreases from 39 to 10 percent 
(bottom panel of Figure \ref{fig:fig1}, right panel of Figure \ref{fig:fig3}, Table \ref{tab:parameters}). Based on these variations of spectral 
and temporal parameters, we understand that this region corresponds to the low hard state 
in the rising phase of the outburst.

We find that the value of photon index is similar to that obtained during 
the 2011 outburst of the source by \citealt{Cap12}. Yet, the hardness ratio is more 
than that in 2011 and is similar to that observed for typical BH 
transients during their LHS \citep{Belloni2005,MR06,Nandi2012}.

\subsubsection{Region BC}

This region consists of observations from days 14.58 to 34.8. During the observations for days 14.58 to 23.0, 
we find that the \emph{XRT} spectral fits can be performed using the model 
\emph{phabs*(powerlaw)}. The $\chi^2/dof$ on day 22.8 is found to be of 
407.84/353. When a disc component is included to this fit the $\chi^2/dof$ is obtained as 407.59/351. 
An F-test to these spectral fits gives the F-statistic value as 0.107 and a probability of 0.89 which is statistically very high. 
The fit results also show the disc temperature value to be 0.794+/-1.44 and disc normalization is 1.330+/-5.85 
(the errors are quoted in 1-$\sigma$) which are statistically incorrect. Thus it is clear that the disc component is not required for these observations. 

An estimate of the hardness ratio shows that it reduces from 0.8 during the LHS to 0.59 during these initial observations. Also, the fractional rms variability decreases from 14\% during LHS to $\approx$ 8\%. Thus we understand that the source has transited from the LHS. 

The \emph{diskbb} model is found to be necessary to the fits from day 24.8 on-wards. 
The spectral fit to the {\it XRT} data on day 34.8 improves the $\chi^2/dof$ from 862.99/433 
to 465.05/431, when a disc component is included along-with the powerlaw. F-test results give 
F-statistic probability of $1.37\times10^{-58}$ implying statistical significance of the disc component.

During the observations in region BC, the photon index is observed to increase 
to 2.0$^{+0.03}_{-0.02}$ as shown in Table \ref{tab:parameters}. The disc temperature $T_{in}$ changes 
randomly between 0.59$\pm{0.05}$ keV and 1.1$\pm{0.05}$ keV (Figure \ref{fig:fig1}). The total flux is observed to increase 
with the maximum being observed on day 25.8 (Figure \ref{fig:fig1}). 
The disc flux is observed to be increasing whereas the powerlaw flux decreases (panels d and e of Figure \ref{fig:fig1}). 

Also from panel g of Figure \ref{fig:fig1} we find that the BAT count 
rate decreases to an average of 0.006. With 
respect to the LHS, the hardness ratio is observed to decrease 
(see region BC in Figure \ref{fig:fig3}). The temporal properties show that the fractional 
rms variability decreases from 7 to 2 percent (panel h of Figure \ref{fig:fig1}, right panel of Figure \ref{fig:fig3}). All these 
characteristics point out that the contribution of hard flux decreases while that of disc 
flux is increasing. Hence we understand that the source is in the HIMS 
for these observations. Variations in the high energy spectra occurring during this HIMS are discussed below in section 4.2.

\subsubsection{Region CD}
\label{simsp}

The observations during days 45.74 to 61.57 have been considered in the region CD in Figure \ref{fig:fig3}. Their energy spectra can be 
well fitted with the model \emph{phabs*(diskbb)}. The powerlaw component is not at all required for the fits. As an example on 
day 56.64 the spectral fit results in a $\chi^2/dof$ of 539/446 when we model it using a disc component. But inclusion of a 
powerlaw component results in incorrect value of the parameters.

During the observations in region CD of the HID, the disc temperature do not vary significantly 
(Figure \ref{fig:fig1}, Table \ref{tab:parameters}). Although the hardness ratio is decreasing, 
it changes randomly between 0.365 to 0.183 (see Figure \ref{fig:fig3}). 
This range of hardness ratio value matches well with 
that observed by \citealt{Cap12} during the intermediate state of 2011 outburst. 

As evident in Table \ref{tab:parameters} the total flux also varies randomly and the BAT count rate decreases to 0.004  
(panels a and f of Figure \ref{fig:fig1} respectively). We note that there is a significant reduction in the fractional rms variability to 4\% along-with the hardness ratio as evident in the HRD (right panel of Figure \ref{fig:fig3}). The fact that the spectra softens in comparison to HIMS and both
hardness ratio and rms value decreases, is strongly indicative of the source being in the SIMS during these observations.

\subsubsection{Point D}
\label{hssp}

On day 63.89 (i.e. MJD 57508.89), we observe that the \emph{XRT} spectrum showed the 
presence of only a disc component. The fit using \textit{model phabs*(diskbb)} results in a $\chi^2/dof$ of 426.16/395. 
When we include a powerlaw the reduced $\chi^2$/dof is found to be of 418.45/393. An ftest to these values show a 
F-statistic of 3.62. Since the value is higher it is evident that the spectra consists of only the disc component. 

The value of hardness ratio estimated after the spectral fit shows that it has 
decreased from those in the other regions and reached to a minimum of 0.159. The flux 
value continues to be closer to the maximum value of 
$\sim 2 \times 10^{-9}$ erg cm$^{-2}$ sec$^{-1}$ (see the asterisk marked in a red diamond 
at point D in both left and right panels of Figure \ref{fig:fig3}; also Table \ref{tab:parameters}). The disc temperature is at its maximum 
of $1.31\pm0.03$ keV. The fractional rms variability is around 3 percent. These variations of the spectral and temporal 
parameters as shown in Table \ref{tab:parameters} indicate further softening of the spectra w.r.t region CD. 

In order to have stronger evidence on this softening, we performed the broad-band spectral fit of the energy spectrum 
(see section 4.2 below for details) corresponding to this observation using \emph{SWIFT-XRT} and \emph{NuSTAR} as shown in 
the Figure \ref{fig:phenoss} (right panel). 
 In this case it is very clear that the high energy spectrum 
is having less counts above 60 keV than observed for HIMS in 
left panel of Figure \ref{fig:phenoss}. It is found that the flux contribution from disc 
is significant (33 percent) in the broad-band spectra (which is more than the contribution 
of 19 percent during the decay phase - see below). The \textit{powerlaw} photon index from the broad-band fit 
is of 2.4 which also indicates a relatively softer state. Also, these variations of spectral and temporal properties of the source 
are similar to those observed for many other black hole binaries like GX 339$-$4, H 1743$-$322 
during their HSS \citep{2001ApJS..132..377H,Belloni2005,MR06,Nandi2012}. Hence, we presume that the source probably has reached HSS around day 63.89.

\subsubsection{Region DE}

After the short duration in HSS, we notice from panel a of Figure \ref{fig:fig1} and 
Figure \ref{fig:fig3} that the source enters its declining phase around 
day 67.02. The observations for the period of 67.02 to 116.0 is represented in region DE.

With respect to point D, we find that on day 67.02 the total flux 
decreases to $1.7\times10^{-9}$ erg cm$^{-2}$ sec$^{-1}$, while the hardness ratio is in the range of 0.183 - 0.269
 (region DE in Figure \ref{fig:fig3}). The spectra are fitted with both disc and 
powerlaw flux models ($\chi^2/dof$ of 324.86/339 on day 79.50). For a few 
of the observations until day 114.80 the spectral fits require only 
powerlaw ($\chi^2/dof$ of 253.77/245 on day 114.80). 

The photon index is observed to vary around 2.2, and disc temperature is observed to now decrease from its 
previous value during the HSS (Figure \ref{fig:fig1}, Table \ref{tab:parameters}). The broad-band 
observation on day 89 (MJD 57534) also give similar value for the photon index and 
disc temperature. The disc emission contributes 19 percent to the total flux which 
is lesser than that found for the broad-band observation in the HSS. 
The temporal studies show that the light curve exhibits variabilities only on day 67.02 
(see last panel of Figure \ref{fig:Variability}), and not for rest of the observations. 
The fractional rms variability is found to be around 6 - 11 percent (panel h of 
Figure \ref{fig:fig1}, right panel of Figure \ref{fig:fig3}) and the PDS have only broad-band noise. These indicate that 
the source has entered SIMS during these observations in the beginning of the decay phase.

\subsubsection{Region EF}

This region corresponds to days 124.9 to 150. For these observations, the spectra are well-fitted with the powerlaw model. 
For the observation on day 114.80 during the decay phase of the source, when the spectra is 
fitted with phabs(powerlaw) it results in a $\chi^2$/dof of 253.77/245. 
Inclusion of a disc component results in a $\chi^2$/dof of 253.27/243. 
An F-test to this results in F-statistic value of 0.239 with probability of 0.787. 
These high values rule out the possibility of presence of a disc component.  

Although there is no disc component and the spectral photon index is observed to be having an average of 2.2 (Figure \ref{fig:fig1}), 
we find that the hardness ratio has increased w.r.t its value during region DE in Figure \ref{fig:fig3}. Also, the total flux 
decreases to $0.7\times10^{-9}$ erg cm$^{-2}$ sec$^{-1}$. The fractional 
rms variability varies between a minimum of 13\% and a maximum of $\sim$22\%, with the PDS exhibiting 
only broad-band noise. The light curve does not exhibit any oscillatory features. The increase in 
hardness ratio, powerlaw flux contribution and the rms w.r.t the region DE suggest that the source is in HIMS of decay phase 
for the observations of region EF. We do not consider these as part of a LHS since the values of photon index, hardness ratio 
and rms variability are lesser in comparison to the following observations discussed below.

\subsubsection{Region FG}

Following the HIMS during days 167 and 172 the spectral fit requires only a powerlaw component. We observe that the photon index of the 
spectra decreases to $<$ 1.7 and the hardness ratio increases to $\sim$0.787${\pm0.073}$ (see Table \ref{tab:parameters}). 
Here, we find that the light curves do
not display any oscillations, and the PDS have only broad-band noise. Since for these observations, the {\it XRT} image displays double streaks, we have excluded the events corresponding to the shortest time interval and hence removed the second streak. We then find that the fractional rms values for these days become 41.43\% and 88.65\% respectively, as shown in
 the panel h of Figure \ref{fig:fig1}. There are no good observations after this day due 
to incorrect imaging in the window-timing mode of {\it XRT} \footnote{Multiple (more than two) streaks seen in the {\it XRT} image probably due to slewing of the SWIFT satellite.} and also lesser counts. So, the observed increase 
in hardness ratio, fractional rms variability and the decline in photon index suggests that the source exists in the hard state of 
decay phase as shown in region FG in Figure \ref{fig:fig3}.

Thus based on the phenomenological spectral modelling, we understand that the source exhibits spectral 
state transitions, which results in a complete `q'-profile for the 2016 outburst. In the following sub-section, 
we present the details of temporal variabilities and QPOs observed during the different HID regions.

\subsection{Presence of coherent variabilities and evolution of QPOs}
\label{coh-qpo}

\begin{figure*} 
	\begin{minipage}{8cm}
	    \includegraphics[height=11cm,width=7.5cm]{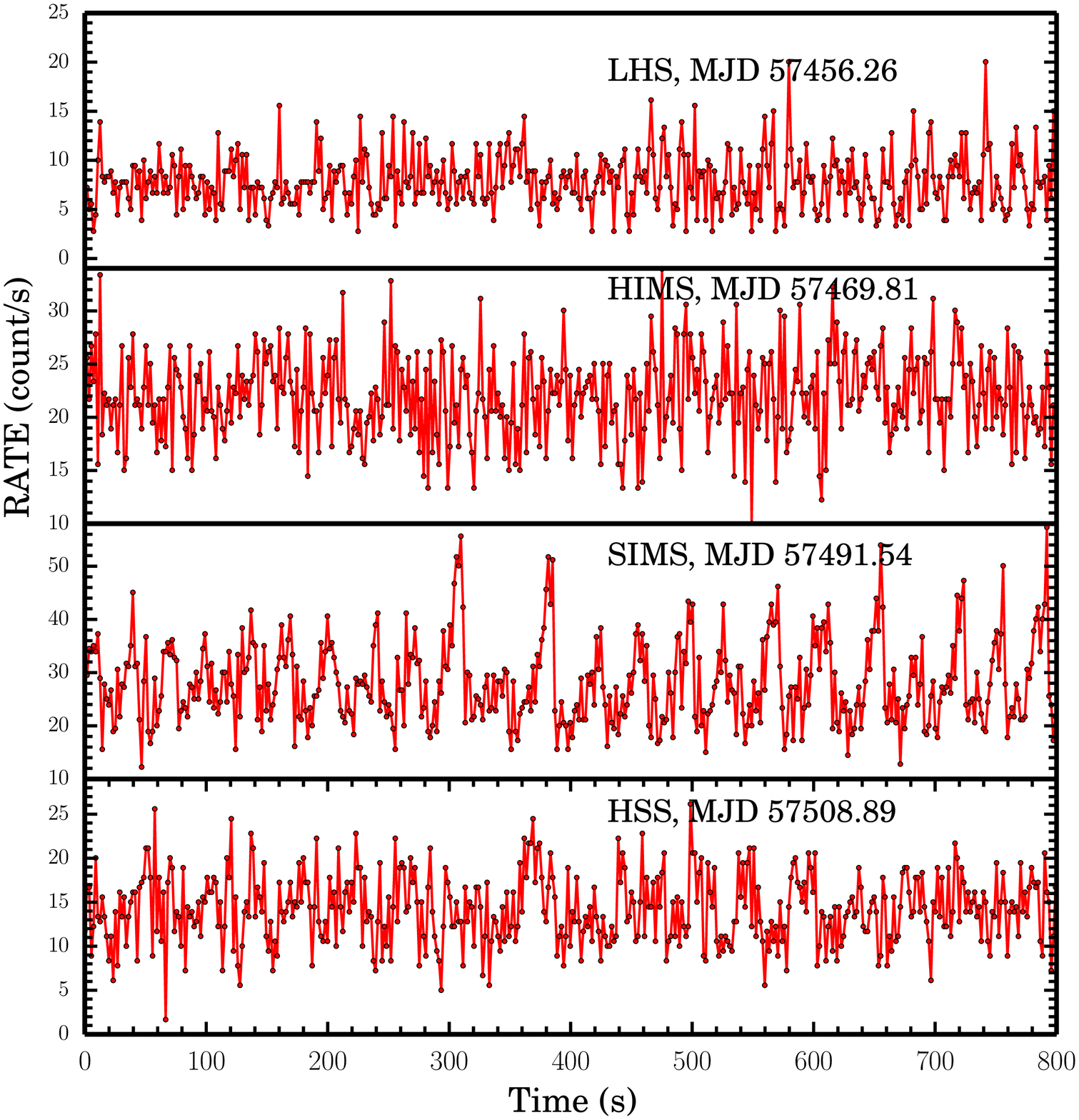} 
	\end{minipage}
	\begin{minipage}{8cm}
	    \includegraphics[height=11cm,width=7.5cm]{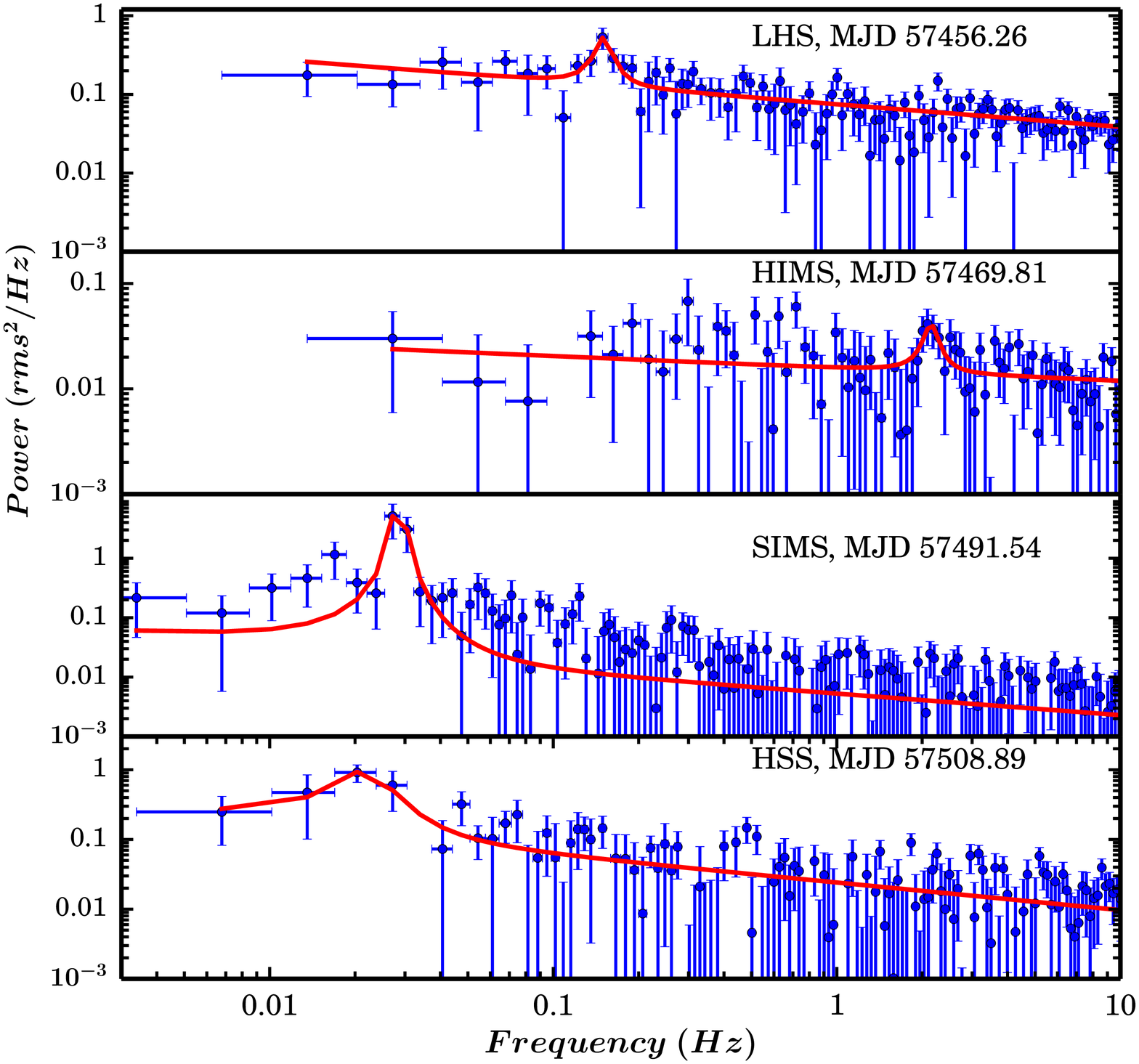}
	\end{minipage}
  \caption{The {\it XRT} light curve and power density spectrum corresponding to the four states 
(LHS, HIMS, SIMS and HSS) of the source IGR J17091$-$3624 during the rising phase of its 
2016 outburst are shown in two panels. The light curves do not show any signature of 
variability during LHS and HIMS. During the SIMS \& HSS variabilities of time 
period 50 to 450 sec are observed. Weak detection of QPOs at 0.15 Hz and 2.15 Hz is seen in 
LHS and HIMS respectively (see Figure \ref{fig:Nustar-PDS} for strong detection of QPOs 
during the same period). Corresponding to the observations where variability is observed, 
QPOs are detected around 20 mHz to 30 mHz during the SIMS and HSS.}
  \label{fig:LCPDS} 
\end{figure*}

For the observations of the source in LHS and HIMS (regions AB and BC), we did not find any signature of 
oscillations/variabilities (left panel of Figure \ref{fig:LCPDS}). 
As the source enters the SIMS (region CD), we observe that the light curves exhibit variabilities/oscillations. This characteristic is similar to 
that observed for this source during its previous outburst in 2011. We find that for 9 continuous 
observations i.e. from days 46.54 to 61.58, these oscillations are present 
(left panel of Figure \ref{fig:LCPDS} and Figure \ref{fig:Variability}). 
The light curves 
have a period of minimum 50 sec on days 49.60, 52.80, 55.26 and a maximum of 450 sec on day 55.26, 
implying the frequency range of tens of mHz. The intensity of the different peaks in the 
light curves varies randomly in the range of 40 to 60 counts/sec. These observations exist in the top left 
portion of the HID (Figure \ref{fig:fig3}) with a clear random variation of both hardness ratio and total flux. 
This variation of source flux in the light curve is similar to that observed during the heartbeat phase of the 2011 outburst where almost
 9 types of variabilities were observed. The source was found to remain trapped in this phase throughout 
the 2011 outburst \citep{Cap12,2017arXiv170309572C}. 
On a comparison with GRS 1915$+$105, the oscillations/variabilities we find for the 2016 outburst can be categorized as similar to the $\rho$-class 
(see \citealt{2001A&A...372..551B} for details on different variability classes).

The temporal analysis for the observation of point D (i.e. a possible HSS) shows that the light curve exhibits 
variabilities/oscillations (see 4th panel on left side of 
Figure \ref{fig:LCPDS}). These oscillations are found to be weaker (maximum intensity of 20 counts/sec) in comparison 
to those observed during SIMS. An energy dependent study of the \emph{NuSTAR} light curve 
shows that the variability is prominent at lower energies. 

\begin{figure}
\includegraphics[width=12cm,angle=-90]{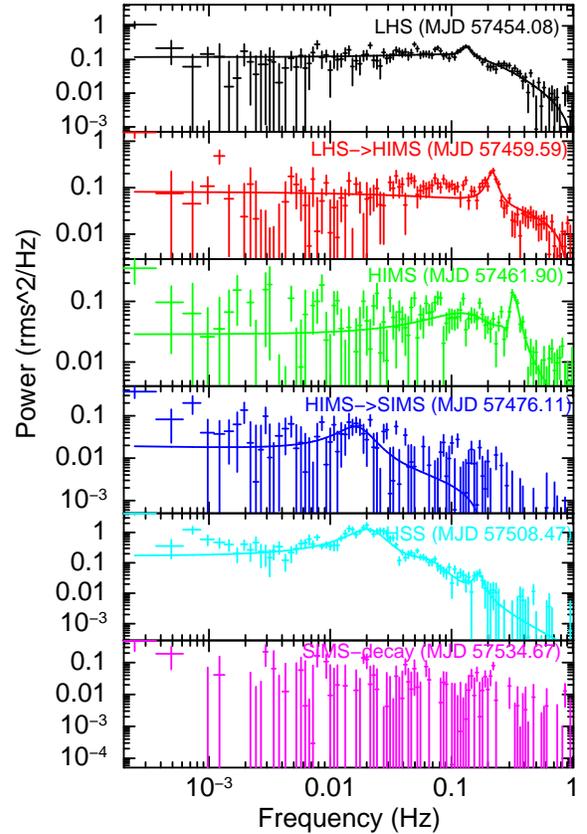}
\caption{Power density spectra during the 6 TOO {\it NuSTAR} observations. The fits to these spectra show clear indication of strong QPOs during LHS, HIMS, transition to SIMS-rise and the possible HSS, while broad-band noise is seen for decay phase of SIMS.}
\label{fig:Nustar-PDS}
\end{figure}

\begin{figure}
\includegraphics[height=8.25cm,width=10cm,angle=-90]{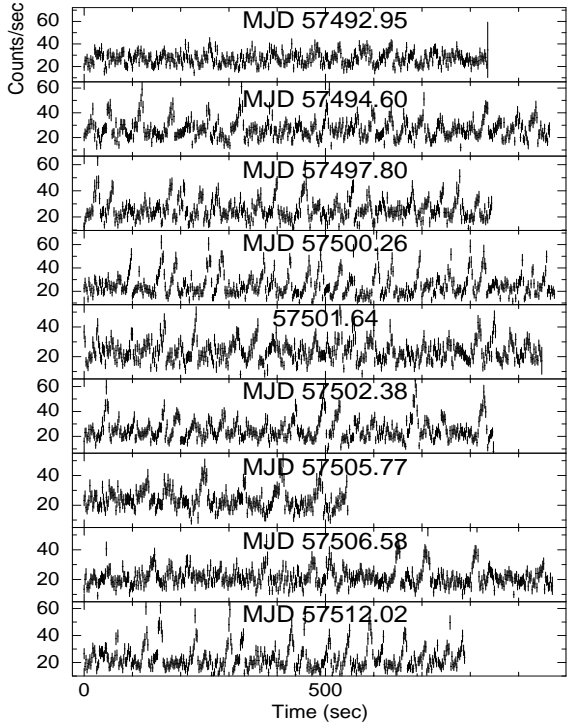}
\caption{Variability/oscillations found during the continuous observations in rising phase of SIMS 
(MJD 57445 corresponds to day 0). The time period varies randomly from 50 to 450 sec for these light curves.}
\label{fig:Variability}
\end{figure}

During the rising phase LHS and HIMS i.e. regions AB and BC respectively, we are able to detect weak QPOs. 
The {\it XRT} PDS show QPOs of frequency increasing from 0.15 Hz to 0.18 Hz 
(see Table \ref{tab:qpo} and right panel of Figure \ref{fig:LCPDS}) during the LHS. A very 
prominent QPO of 0.13$\pm{0.03}$ Hz with rms amplitude of 6.20 percent is seen 
in the \emph{NuSTAR} PDS during day 9 (top panel of Figure \ref{fig:Nustar-PDS}; see also \citealt{Xu17}). 
Previous outburst of this source has shown presence of constant frequency QPOs 
during its LHS \citep{Iyer13,2015ApJ...807..108I}.

The QPO frequency increases up to a maximum of 2.15 Hz during the HIMS. The quasi-simultaneous \emph{NuSTAR} 
observations during this period also indicate strong signatures of QPOs of 
frequencies $0.217\pm0.002$ Hz on day 14.59 (see \citealt{Xu17}) and $0.326\pm0.005$ on day 16.90 respectively 
(see the 2nd and 3rd panels of Figure \ref{fig:Nustar-PDS}). 
This is similar to the 2011 outburst where also a gradual increase of QPO 
frequencies was observed \citep{Iyer13,2015ApJ...807..108I}. 

Based on the frequency range, their rms and Q-factor, these QPOs observed during hard and hard-intermediate states can be considered of type C. These QPOs are found to evolve with time i.e. the QPO frequency increases as the source moves from hard to hard-intermediate state. In our recent paper \citealt{Sree2018} we we have studied this evolution and discussed the physical significance of the same. 

We find that just a few days 
before the transition to the next state i.e. on day 31.11 (MJD 57476.11) the \emph{NuSTAR} 
light curve shows weak signature of variabilities. The resultant PDS gives a 
weak 16.4$\pm$2 mHz QPO (see 4th panel of Figure \ref{fig:Nustar-PDS} and 
Table \ref{tab:qpo}). 

A detailed study of the PDS for the 2016 outburst show that for all the observations during the SIMS, the power spectrum has a powerlaw nature. These power spectra give strong indication of the presence of very low frequency QPOs 
(see right panel of Figure \ref{fig:LCPDS}) in the range of 20 to 30 mHz. It has to be noted that a few of the QPOs have 
lesser values of Q-factor and amplitude unlike the typical values found in other BH sources \citep{Casella2004}. The values of 
different parameters of these QPOs are given in Table \ref{tab:qpo}.

During the observation corresponding to point D of HID, both the \emph{XRT} 
and \emph{NuSTAR} PDS indicate presence of a low frequency broad QPO like feature 
at 21.4 mHz and 20.7 mHz respectively. The latter has a strong significance 5.32 
and rms amplitude of 5.71 percent (see Table \ref{tab:qpo} and 5th panel of 
Figure \ref{fig:Nustar-PDS}). In the {\it NuSTAR} PDS there is also a weak peaked component 
around 0.16 Hz which has a Q-factor of 5, rms amplitude of 1.58 but significance 
of 2.3 only. An energy dependent study of the \emph{NuSTAR} observation 
shows that the broad mHz QPO and weak feature are observed only at lower
energies (3 - 25 keV).

\begin{table*}
\caption{Values of QPO frequencies observed during the rising phase of LHS, HIMS, SIMS and `possible' HSS for IGR J17091$-$3624. mHz QPOS 
are observed during the days when the light curve exhibits oscillations. Strong presence of QPOs is indicated by the \emph{NuSTAR} observations 
during the different spectral states. Here, MJD 57445 has been considered as day 0.}
\begin{tabular}{ccclccc}
\hline
MJD & Day & Instrument & QPO frequency & Q-factor & Significance & amplitude\\
& & & (Hz) & & &  (rms \%)\\
\hline
LHS & \\
57454.08 & 9.08 & {\it NuSTAR} & 0.133$\pm{0.03}$ & 6.04 & 4.52 & 6.20  \\
57456.26 & 11.26 & {\it XRT} & 0.151$\pm{0.01}$ & 6.55 & 1.68 & 4.36\\
\hline
HIMS & \\
57459.58 & 14.58 & {\it XRT} & 0.189$\pm{0.01}$ & 4.25 & 1.32 & 3.79\\
57459.59 & 14.59 & {\it NuSTAR} & 0.217$\pm0.002$ & 7.35 & 6.45 & 1.88   \\\\
57461.80 & 16.80 & {\it XRT} & 0.332$\pm0.005$ & 7.81 & 1.30 & 2.77 \\
57461.90 & 16.90 & {\it NuSTAR} & 0.326$\pm0.005$ & 5.82 & 5.75 & 1.89 \\\\
57469.81 & 24.81 & {\it XRT} & 2.149$\pm0.1$ & 7.16 & 1.8 & 2.58\\
\hline
HIMS$\rightarrow$SIMS&\\
57476.11 & 31.11 & {\it NuSTAR} & 16.4${\pm 2}$ mHz & 1.31 & 3.92 & 0.61  \\
\hline
SIMS & \\
57491.54 & 46.54 & {\it XRT} & 28$^{+1}_{-2}$ mHz & 3.91 & 3.81 & 3.71 \\
57494.61 & 49.61 & {\it XRT} & 25${\pm 2}$ mHz & 1.15 & 4.81 & 4.30\\
57497.80 & 52.80 & {\it XRT} & 36$^{+2.4}_{-3.2}$ mHz & 4.17 & 4.22 & 8.92\\
57500.26 & 55.26 & {\it XRT} & 28${\pm 3}$ mHz & 2.07 & 2.95 & 4.81 \\
57502.38 & 57.38 & {\it XRT} & 28${\pm 2}$ mHz & 3.57 & 3.35 & 3.87 \\
57505.77 & 60.77 & {\it XRT} & 30${\pm 1.5}$ mHz & 1.76 & 6.50 & 5.90\\
57506.58 & 61.58 & {\it XRT} & 19.7$^{+4}_{-3}$ mHz & 1.32 & 2.52 & 3.71\\
\hline
HSS & \\
57508.47 & 63.47 & {\it NuSTAR} & 20.7${\pm 0.06}$ mHz & 0.92 & 5.32 & 5.71\\
57508.89 & 63.89 & {\it XRT} & 21.4${\pm0.001}$ mHz & 1.69 & 3.50 & 3.56\\
\hline\\
\end{tabular}
\label{tab:qpo}
\end{table*}

In the next subsection we explain the results obtained by modelling the broad-band spectra ({\it XRT + NuSTAR}) using the phenomenological models. 

\subsection{Phenomenological modelling of the broadband spectra}
As mentioned earlier for the 2016 outburst of IGR J17091-3624 there are six {\it NuSTAR} observations of which four are quasi-simultaneous observations with \textit{SWIFT-XRT}. We fit all these broadband spectra with different models to search for the presence of various features like high energy cut-off, reflection and presence of disc. 

It has to be noted that although the {\it NuSTAR} exposure time is longer, the duration of {\it XRT} observation falls within the {\it NuSTAR} exposure period. These are not strictly simultaneous, but can be called quasi-simultaneous. The black hole accretion systems are generally dynamic and significant spectral changes may occur over the duration. Any considerable change in the spectrum in this period would have been evident from the energy spectrum obtained from the two instruments. But we observe that there is good overlap between the spectra from two instruments with no evident offset. We also note that the combined spectral fit of the quasi-simultaneous observations results in $\chi^2$/dof value around unity.

We initially modelled all these observations with \textit{compTT} and the $\chi^2$/dof for these cases are provided in the appendix in Table \ref{tab:bbfit}. Although the $\chi^2$/dof values look reasonable statistically, there are lots of internal inconsistencies in the model parameters. For example on day 63 (MJD 57508), the model \textit{diskbb+compTT} fits the data well but the disc temperature Tin = 1.097 keV is not consistent with soft photon temperature T$_{0}$ = 0.08 keV of \textit{compTT}.  In fact extra disc component should not have been required along with \emph{compTT}, but for a good statistical fit \emph{diskbb} is required. Hence we do not present here the parameter studies based on \textit{compTT} model.

So we took the broadband spectra corresponding to HIMS and fitted it with a \textit{powerlaw} component. This resulted in a $\chi ^2/dof~=~1475.41/1116$. The fit improved when we used a \textit{cutoffpl} instead of the \textit{powerlaw} giving a $\chi ^2/dof~=~1290.67/1115$. This fit had some large residual values around 30 keV. We added a component \textit{ireflect} to take care of the reflection contribution from ionized material and this reduced the $\chi ^2/dof$ to $1202.27/1113$. So we could obtain a decent $\chi^2$/dof value with the model \textit{phabs(ireflect*cutoffpl)} for the case of hard and hard-intermediate states. But for fitting the spectra of the softer states the inclusion of the \textit{diskbb} component is required. Thus we use the most general phenomenological model containing a Keplerian disc, a power-law with high energy cut-off and a component to take care of reflection from ionized material. The model we have chosen is \textit{phabs(diskbb + ireflect*cutoffpl)}. 
The results of all these modelling are provided in the Appendix in Table \ref{tab:bbfit}. 
We present here the evolution of spectral parameters during the different spectral states as the source evolves in the outburst. 

\begin{figure*}
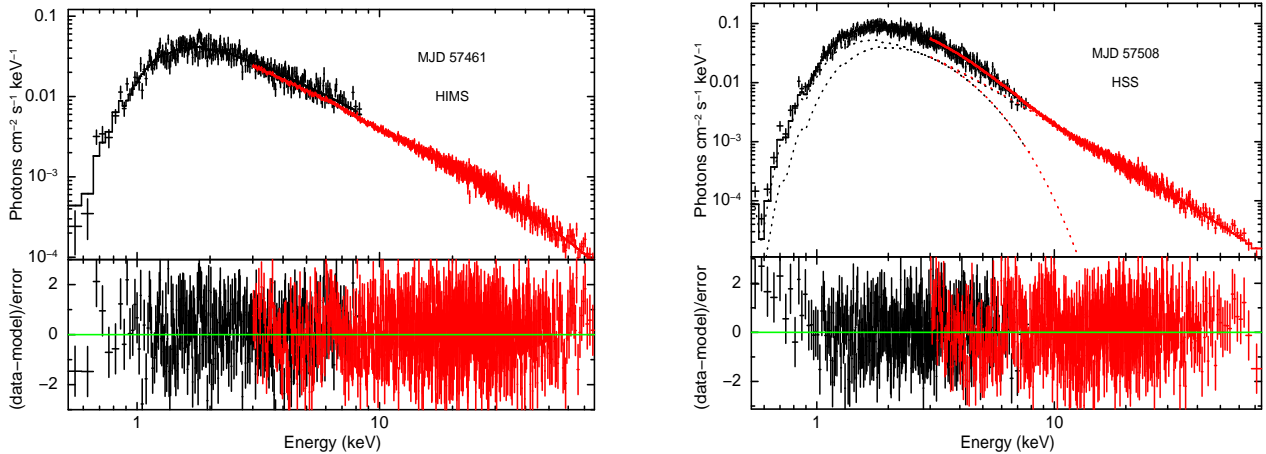
 
  \begin{minipage}{9cm}
	\includegraphics[height=8.25cm,width=6cm,angle=-90]{figure4a.eps}
  \end{minipage}
  \begin{minipage}{8cm}
	\includegraphics[height=8cm,width=6cm,angle=-90]{figure4b.eps}
  \end{minipage}
  \caption{Spectral fitting performed for broad-band (0.5 - 79 keV) {\it XRT} and \emph{NuSTAR} data 
on day 16.9 (MJD 57461.9) and 63.89 (MJD 57508.89) are shown in the left and right panels 
respectively. The spectral model of \emph{phabs*(ireflect*cutoffpl)} is required for a 
statistically good fit for left panel, while the right panel spectrum is modelled by 
using \emph{phabs*(diskbb+ireflect*cutoffpl)}. See text for details.}
  \label{fig:phenoss} 
\end{figure*}

The first three observations (i.e. days 9, 14, 16) suggest that (see Table \ref{tab:phenofit}) a high energy cut-off is required for the spectral fits during the rising phase of the outburst. The value of cut-off indicates a decreasing trend from $228^{+72}_{-44}$ keV to $98^{+12}_{-9}$ keV as the source transits from LHS to HIMS. The left panel of Figure \ref{fig:phenoss} shows the phenomenological modelling of the broadband observation on day 16 (MJD 57461). The spectra on day 89 (MJD 57534; SIMS decay) also shows a weak signature of a cut-off. It should be noted that the \emph{cutoffpl}
is given by $A(E)=KE^{-\alpha}exp(-E/\beta)$ where $\alpha$ is \emph{powerlaw} photon index, $\beta$ is the high energy cut-off in keV and K is the normalization. There is no restriction on the value of
$\beta$. At $\beta=\infty$ we get back the normal \emph{powerlaw}, $A(E)=KE^{-\alpha}$ . From Table \ref{tab:bbfit} it is clear that the improvement in $\chi^2$/dof is significantly larger in HIMS than in LHS when we use a \emph{cutoffpl}. This is because a larger high energy cut-off value results in a smaller deviation from the normal \emph{powerlaw} within the instrument energy range.

Table \ref{tab:phenofit} shows that a reflection component is required for the observations on days 9 (MJD 57454), 14 (MJD 57459), 16 (MJD 57461) and 31 (MJD 57476).
Besides this we see a trend in the evolution of photon index from 1.58$\pm{0.028}$ in the LHS to 2.41$\pm{0.01}$ in the HSS. And then the photon index reduces to 2.03$\pm{0.032}$ as the source enters the decay phase SIMS. Thus the state evolution is quite obvious from the broadband observations itself.
It is also observed that the presence of disc component has begun from rising phase SIMS on day 31 (MJD 57476) at a \textit{Tin} of 0.59 keV and it increased to 1.06 keV during the HSS. The phenomenological modelling corresponding to HSS is shown in the right panel of Figure \ref{fig:phenoss}. The presence of strong thermal emission during the HSS is evident from this figure as discussed earlier. In the next sub section we use a spectral model based on the two-component accretion flow to study the spectra and estimate accretion parameters for each state of the system.  

\begin{table*}
    \centering
    \caption{Broadband phenomenological fit parameters using the model \textit{phabs(diskbb+ireflect*cutoffpl)}}
    \label{tab:phenofit}
    \begin{tabular}{lccccccr} 
        \hline
        MJD & Observatory&Tin & $rel\_refl$ & Photon Index&high-cut (keV)&$\chi ^2$/dof\\
        \hline
        57454 (LHS)&{\it NuSTAR}&$-$ &$0.39^{+0.06}_{-0.06}$&$1.58^{+0.028}_{-0.028}$&$228^{+72}_{-44}$&1011.08/986=1.02\\
        57459 (LHS$>$HIMS)&{\it NuSTAR+XRT} & $-$ &$0.27^{+0.07}_{-0.06}$&$1.55^{+0.02}_{-0.02}$&$117^{+18}_{-14}$&1234.81/1218=1.01\\
        57461 (HIMS)&{\it NuSTAR+XRT}&$-$ &$0.34^{+0.07}_{-0.07}$&$1.58^{+0.020}_{-0.019}$&$98^{+12}_{-09}$&1202.27/1113=1.08\\
        57476 (SIMS)&{\it NuSTAR}&$0.59^{+0.014}_{-0.015}$ &$0.60^{+0.07}_{-0.08}$&$2.41^{+0.017}_{-0.020}$&$-$&909.25/811=1.12\\
        57508 (HSS)&{\it NuSTAR+XRT}& $1.06^{+0.006}_{-0.006}$ &$-$&$2.41^{+0.01}_{-0.01}$&$-$&967.07/966=1.00\\
        57534 (SIMS decay)&{\it NuSTAR+XRT}& $1.06^{+0.022}_{-0.024}$ &$-$&$2.03^{+0.032}_{-0.032}$&$262^{+58}_{-40}$&970.46/943=1.02\\
        \hline
	\end{tabular}
\end{table*}

\subsection{Modeling with two component accretion flow}

\subsubsection{Model description}
\label{model}

In this paper, we attempt to model the energy spectra of 2016 outburst of the source using the two component accretion flow
 paradigm \citep{ST95,2005A&A...434..839M,2015ApJ...807..108I}. The model considers 
two components - a Keplerian disc 
\citep{1973A&A....24..337S} at the equatorial plane which produces the soft photons, and a sub-Keplerian
 halo on top and bottom of the Keplerian disc. 
The sub-Keplerian flow close to the central object creates an effective boundary in presence of shocks \citep{1996ApJ...464..664C,2004MNRAS.349..649C,2007MNRAS.376.1659D,2011IJMPD..20.1597C}. Or else it forms a pileup of matter due to centrifugal barrier. This central region 
acts as the Compton corona/clouds, which inverse-Comptonizes the soft photons producing the high energy emission.
In this model the total radiation spectrum
is calculated self-consistently from hydrodynamics. Therefore, the low energy soft photons from Keplerian disc and 
Comptonized components (high energy) are not independent like phenomenological models ($diskbb+powerlaw$). 

The two component flow model consists of four parameters, namely, shock location ($r_s$ in units of $r_g=2GM/c^2$) which represents 
the size of the Compton corona, Keplerian disc rate ($\dot{m}_d$) and sub-Keplerian halo accretion rate ($\dot{m}_h $) in units of $\dot{m}_{Edd}$
and mass (M) of the BH in units of solar mass M$_{\sun}$. 
Earlier, we have imported the radiation spectra generated from two component advective model as a local additive 
model into \texttt{XSpec} using $atable$ command and fitted the broad-band spectral data in the range of 0.5 - 100 keV 
(see \citealt{2015ApJ...807..108I} and references therein for details) for the 2011 outburst of IGR J17091$-$3624.
Following similar methodology, here, we perform broad-band spectral modelling (0.5 - 79 keV) using \texttt{XSpec}, of four
quasi-simultaneous observations during the 2016 outburst of IGR J17091$-$3624 using \textit{Swift} and \textit{NuSTAR} data.
In order to get error estimates for the parameters of the table model, we use the Migrad method.

Once we obtain satisfactory fits and error ranges, we compute the unabsorbed flux values corresponding to the model
using $cflux$ command.  
We estimate the unabsorbed fluxes in two different energy ranges 0.5 - 4 keV and 4 - 10 keV, for all 51 \textit{SWIFT-XRT} 
observations modelled with two component flow. Then we take the ratio of flux in 4 - 10 keV to flux in 
0.5 - 4 keV to obtain the hardness ratio. The plot of total flux versus hardness ratio for two component flow is shown in Figure \ref{fig:Qplot} in red stars. On the same figure we over-plot the total flux versus hardness ratio as obtained from the phenomenological model in blue stars for the sake of comparison.

According to two component flow model, a source enters into outburst with significantly larger halo accretion rate 
than the Keplerian disc accretion rate. 
As the outburst progresses in the rising phase, the Keplerian disc starts contributing to 
the source luminosity and sub-Keplerian matter becomes less important. 
In the dynamical process of evolution, the corona cools down and shrinks in size as the outburst progresses. 
During the peak of the outburst, the Keplerian disc is prominent and hence the disc accretion rate is 
higher than before. A reverse trend occurs in the declining phase of the outburst.

Below we summarize the modelling of the broad-band observations using this two component flow model, and evolution of the different parameters. We also give an account of the source mass estimation performed using the same model.

\begin{figure*}
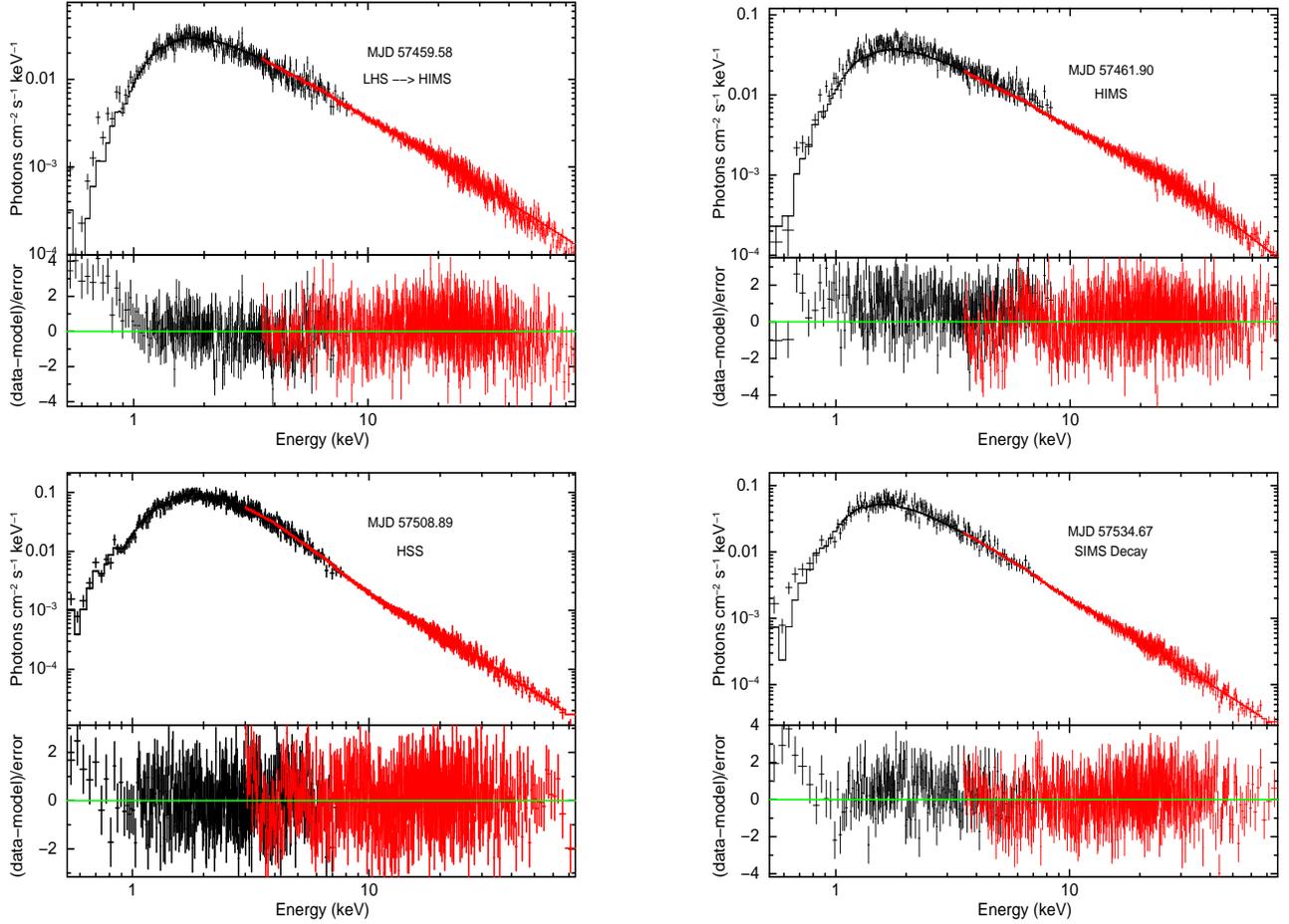

\begin{minipage}{9cm}
\flushleft
\includegraphics[height=8cm,width=6cm,angle=-90]{figure11a.eps}
\end{minipage}
\hskip 0.1in
\begin{minipage}{8cm}
\includegraphics[height=8cm,width=6cm,angle=-90]{figure11b.eps}
\end{minipage}
\vskip 0.1in
\begin{minipage}{9cm}
\includegraphics[height=8cm,width=6cm,angle=-90]{figure11c.eps}
\end{minipage}
\hskip 0.1in
\begin{minipage}{8cm}
\includegraphics[height=8cm,width=6cm,angle=-90]{figure11d.eps}
\end{minipage}
\caption{Modeling of broad-band energy spectra (0.5 - 79 keV) using the two component advective flow model for
different spectral states in the order of LHS to HIMS transition, HIMS, HSS for rising phase and SIMS of 
decay phase. \textit{Swift-XRT} data is shown with black points, and
\textit{NuSTAR} data is shown with red points. See text for details.}
\label{fig:Spec_states}
\end{figure*}

\subsubsection{Broad-band energy spectra and mass estimation} \label{ss:bbspec}
We model all six \emph{NuSTAR} observations with the two component accretion flow model.
Four of these are broadband observations were we have quasi-simultaneous \emph{XRT} observations also. 
An additional model component $highecut$ is required only during the HIMS in-order to model 
the high energy cutoff with cut-off energy 121 keV. 
The requirement of a high energy cut-off is related to the way the two component flow model addresses the reflection contribution. 
In this model, the reflection contribution is calculated \citep{ST95} based on analytical solution of Fokker-Planck equation 
(following \citealt{1975lpsa.book.....S}) which gives asymptotic
results. A Montecarlo calculation is required to address more general situations. The reflection estimation in two component model
works fine when the reflection contribution is moderate. So here we find that the model is able to fit all the data sets except one data set. 
For this particular data set a strong reflection component is required and hence the required e-folding energy is lower than the same
calculated from the model. Hence we require an additional high energy cutoff component while performing the analysis. Usually we find that a {\it smedge} component alone is required for the fits along-with the two component flow model.

For the spectral fitting we use a fixed n$_H$ value of $1.1 \times 10^{22}$ atoms cm$^{-2}$. For the HSS we required the introduction of an additional \emph{pcfabs} component that takes care of the partial covering fraction absorption. The covering fraction for this observation is 0.87 and \emph{pcfabs} has an n$_H$ value of $1.16 \times 10^{22}$ atoms cm$^{-2}$. This indicates the presence of winds in the system \citep{Sav12,Koljo13,RNVS16b,Pahari17}. 
The model fitted broad-band spectra for different states (see section \ref{sptmpe} and 
yellow stars in Figure \ref{fig:fig3}) during the outburst evolution are shown in Figure \ref{fig:Spec_states}. 
In this figure, \textit{SWIFT} data is plotted in black and \textit{NuSTAR} data
is plotted in red color.
We then estimate the model parameters with $90\%$ confidence using Migrad method as mentioned earlier and 
the results are summarized in Table \ref{tab:resfit}.

\begin{table*}
    \centering
    \caption{Broadband fit parameters and chi squared values obtained using two component flow model.}
    \label{tab:resfit}
    \begin{tabular}{lccccr} 
        \hline
        MJD & Observatory&Mass ($M_{\odot})$ & halo rate $(\dot{M} _{Edd})$ &disk rate $(\dot{M} _{Edd})$&$\chi ^2$/dof\\
        \hline
        57454 (LHS)&{\it NuSTAR}&$11.66 \pm 0.12$ & $0.36\pm 0.002$ &$0.04\pm 0.0006$ &1183.82/987=1.19\\\\
        57459 (LHS$>$HIMS)&{\it NuSTAR+XRT} & $11.27 \pm  0.21$ &$0.31\pm 0.0007$ &$0.04\pm0.001$&1278.28/1089=1.17\\\\
        57461 (HIMS)&{\it NuSTAR+XRT}& $11.70 \pm  0.13$  &$0.30 \pm 0.005$ &$0.04\pm 0.001$&1253.43/1100=1.13\\\\
        57476 (SIMS)&{\it NuSTAR}&$11.51 \pm 0.07$ &$0.10 \pm 0.0007$ &$0.31 \pm 0.009$&1037.44/810=1.28\\\\
        57508 (HSS)&{\it NuSTAR+XRT}& $11.12 \pm  0.10$  &$0.08\pm 0.0004$ &$0.39\pm 0.010$&1042.14/963=1.08\\\\
        57534 (SIMS decay)&{\it NuSTAR+XRT}& $10.93 \pm  0.10$  &$0.10 \pm 0.0004 $&$0.32\pm 0.004$&950.24/855=1.11\\\\
        \hline
	\end{tabular}
\end{table*}

In two-component model, the mass and the accretion rates of the source
self-consistently determine the density and temperature distribution of
the flow. This in turn estimates the spectral signatures like fraction of inverse-Comptonized
black body photons, energy spectral index etc. So, the model must choose the
correct mass of the source to match all the spectral features found in different observed
data sets. The model normalization is just a constant scaling factor.
The overall normalization, $N \propto cos(i)/D^2$, depends on the inclination angle $(i)$ of 
the disc normal to the observer and the distance (D expressed in unit of 10 kpc) to the source. 
We can not fix the value of $N$ since both $i$ and D are unknown. But we have to choose a constant $N$ 
if we wish to estimate the mass of the source consistently across all observations.

It has to be specifically noted that we do not assume any particular value of $D$ or $i$ in 
our calculation. We have rather adopted the following procedure to fix the value of $N$.
First, we fit all the broadband data sets keeping $N$ as a free parameter until we get the best fit. From this we
 find the range $N_{min} < N < N_{max}$, where $N_{min}=1.5$, $N_{max}=4.27$ corresponds to different data sets. 
Then we refit all the four broadband spectra with 
normalization frozen to the mean value $(N_{av}=2.885)$ obtained from these four observations 
and use this result to estimate the mass of the source. 
Normalization, in principle, can be any constant value between $N_{min} -  N_{max}$ and 
different values of $N$ in the range can cause variations in the estimated mass. We have 
taken this systematic effect in our calculation as well.

\begin{figure}
\includegraphics[height=6.7cm,width=8cm]{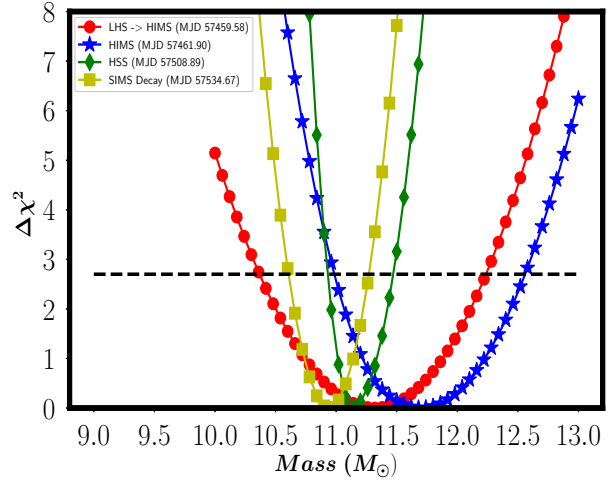}
\caption{ $\Delta \chi^2$ versus mass of the source during the different 
spectral states obtained from $steppar$ command output in \texttt{XSpec} are shown for the mean norm value of 2.885 (see text for details)}.
\label{fig:Del_mass}
\end{figure}

\begin{figure}
\includegraphics[height=6.7cm,width=8cm]{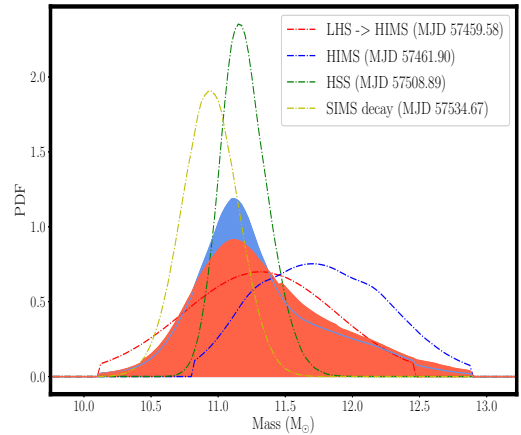}
\caption{PDFs obtained from $\Delta \chi^2$ confidence curves for four broadband spectra. 
	The blue shaded curve with ``SUM PDF'' represents the combined PDF by summing all 
	four PDFs with $N_{av}$ whereas the 
	red shaded curve shows the combined PDF obtained by considering various
	normalizations between $N_{min}$ and $N_{max}$. The 90\% mass range is between $10.62 - 12.33 M_\odot$.}
\label{fig:pdf}
\end{figure}

Once we obtain satisfactory fits for the all four broad-band energy spectra with $N=N_{av}$, the \emph{steppar}
command is used to obtain the $\Delta \chi^2$ values as a function of mass parameter in
all four cases. This is plotted in Figure \ref{fig:Del_mass}, where red, blue, green and yellow curves 
represent LHS, HIMS, HSS and SIMS (decay) spectral states respectively.
Then, we  convert the confidence intervals (as obtained by \emph{steppar}) to
probability distribution functions (PDF) using the steps followed in
\citet{2015ApJ...807..108I}. The probability distribution functions are plotted in Figure \ref{fig:pdf} 
for all four observations with the same colour references as in \ref{fig:Del_mass}.
The PDFs are then combined (shown as ``SUM PDF'' in Fig. \ref{fig:pdf}) by summing the four PDFs
to obtain the lower and upper limits on mass from the four observations.

In order to estimate the effect of arbitrariness in the normalization values on mass, the process mentioned above for
$N_{av}$ is repeated for the different values of $N$ in the range $N_{min} - N_{max}$. Then, we get the combined 
PDF in this case (shown as ``SUM PDF (with systematics)'' in Fig. \ref{fig:pdf}) by summing all PDFs having different 
constant $N$ values. We see that it does not differ much from the combined PDF
(blue shaded curve) for $N_{av}$.
Hence it shows that the arbitrariness of the normalization value on the estimated mass is minimal. 
The mass range for 90\% confidence level is between 10.65 - 12.24 M$_\odot$ for average norm. The same including 
systematic lies in the range 10.62 - 12.33 M$_\odot$.
It is noted that in this method the mass of the source can be directly estimated from the fitting process as an 
independent fit parameter. Also the estimated mass is not very much dependent on the
normalization, inclination angle and distance to the source.

\subsubsection{Evolution of model parameters and `q'-profile}

\begin{figure}[h]
\includegraphics[height=7cm,width=8.4cm]{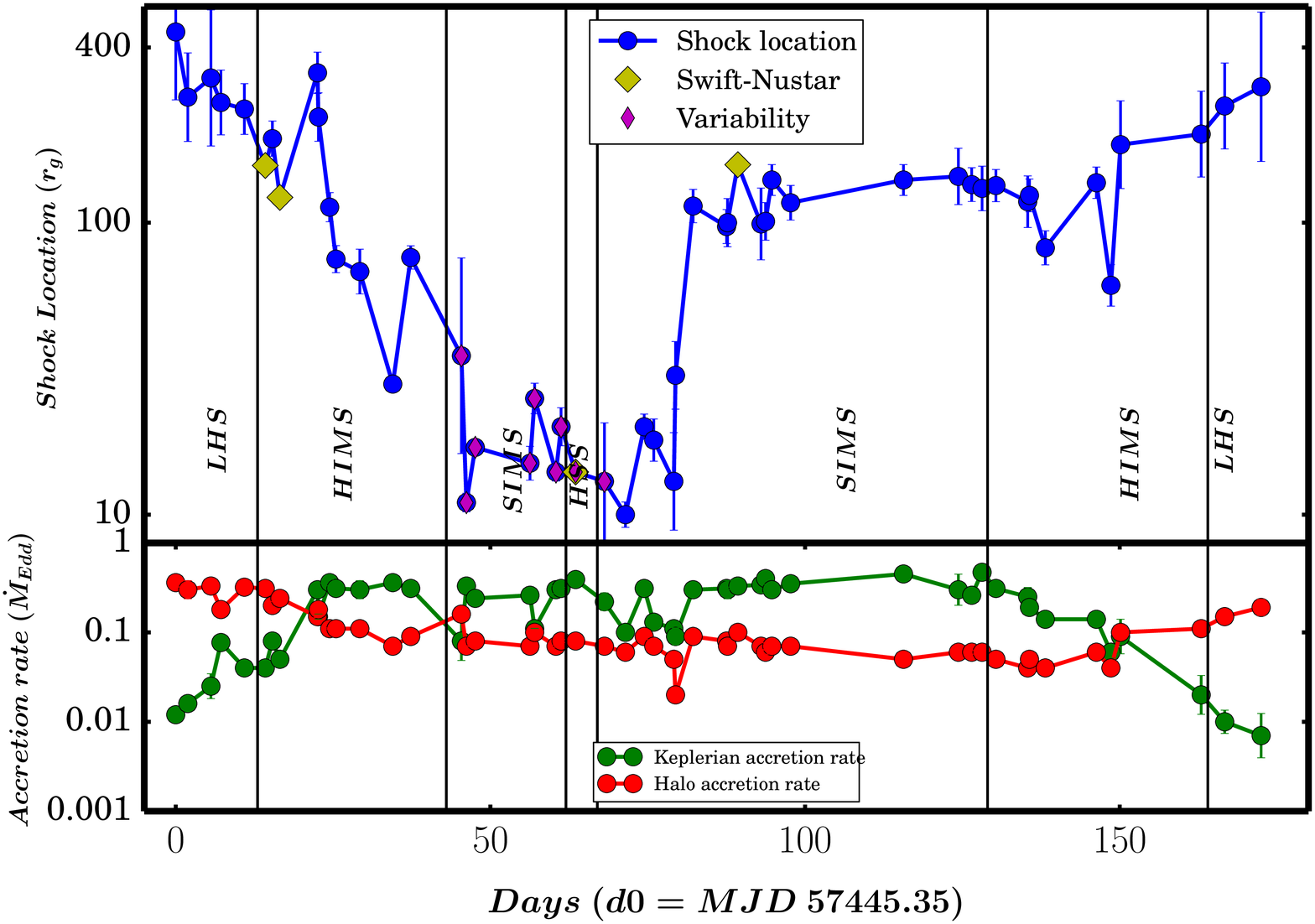}
\caption{Variation of shock location and accretion rates with time are plotted. The shock location decreases as the
source transits from LHS to HSS and vice-versa. The halo accretion rate decreases whereas Keplerian accretion rate
increases during rising phase and becomes comparable in the intermediate states before and after the HSS. In the
decay phase both accretion rate show reverse behaviour. Data points corresponding to the four broad-band spectra 
that we have presented before are marked with yellow diamonds. The days when variability in SIMS is seen are marked with magenta diamonds.} 
\label{fig:Accrn}
\end{figure}

\begin{figure}[h]
\includegraphics[height=9cm,width=8.25cm]{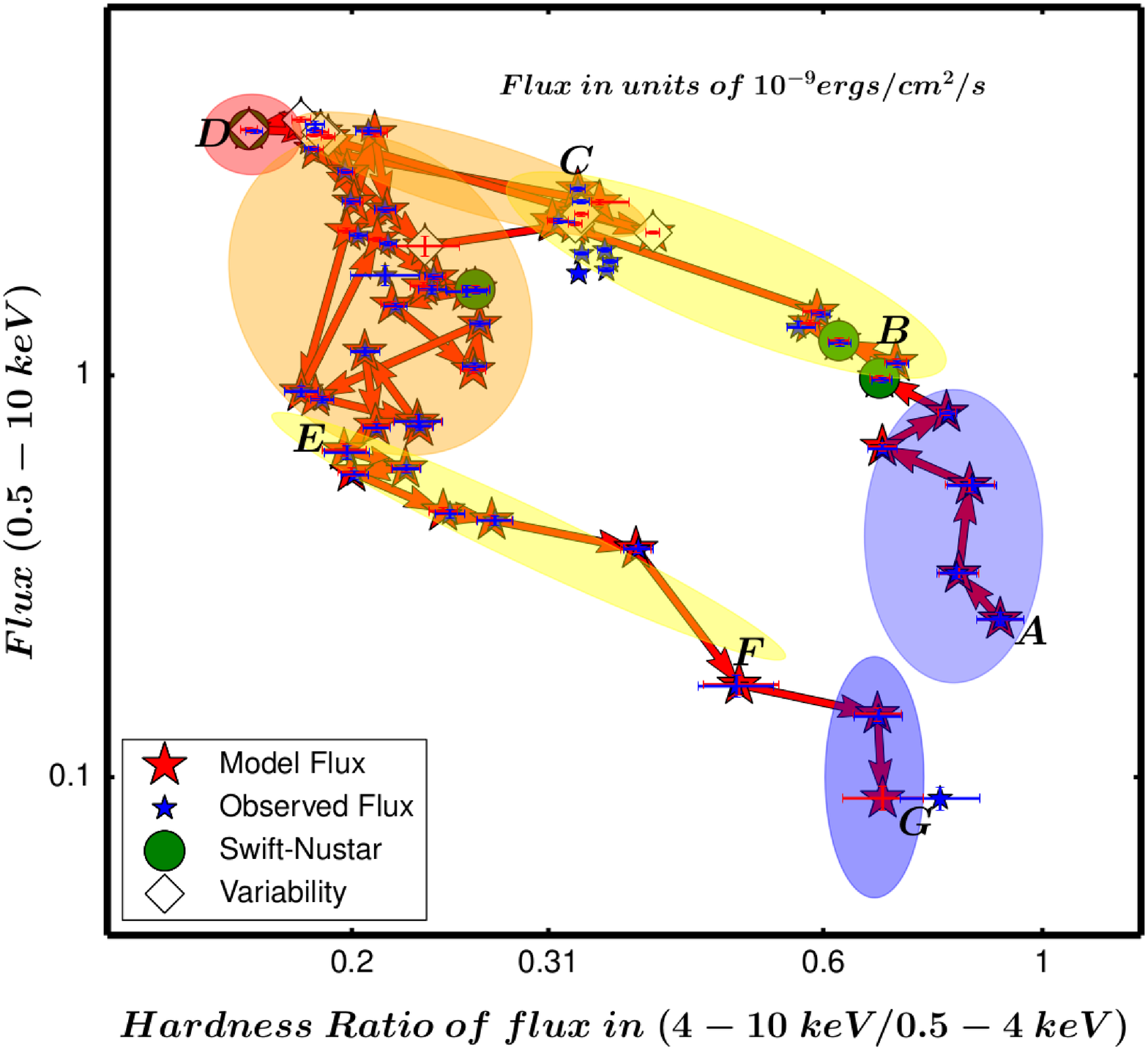}
\caption{HID showing the variation of total flux versus the ratio of flux in 4 - 10 keV to  
0.5 - 4 keV obtained with two component advective model fitting. The LHS, HIMS, SIMS and HSS of rising and 
decay phase are shown with blue, yellow, orange and red patches respectively. See text for details.}
\label{fig:Qplot}
\end{figure}

In Figure \ref{fig:Accrn}, we show the variation of shock location (size of Compton cloud) and accretion 
rates (Keplerian and sub-Keplerian). The state transitions are marked with
vertical lines during the 2016 outburst. Figure \ref{fig:Qplot} represents
the evolution
of total flux in the range of 0.5 - 10 keV with respect to the hardness ratio based on two component flow
model. Below we give the values of different parameters of our physical model. The units for each of these parameters are given in section \ref{model}.

During the initial few days of the outburst, we find that the source is
in the LHS as shown in blue patch in Figure \ref{fig:Qplot}. In the LHS the
shock location varies from $r_s=450$ to $r_s = 157$ and the halo rate is at its
highest with a value of $\dot{m}_h = 0.36$. Whereas the disk rate
in LHS is only $\dot{m}_d = 0.01 - 0.07$.  As the system
enters the HIMS (identified by yellow patch in Figure \ref{fig:Qplot}) the disk
rate increases and exceeds the halo rate and the shock location gradually
decreases. Here the Keplerian accretion rate reaches up to an average value of
$\dot{m}_d = 0.25$ and $\dot{m}_h$ ranges from $0.20$ to $0.09$.
 
After the HIMS the state of the system changes to SIMS where the
shock location suddenly decreases and settles to a mean value $r_s=20$. This
indicates that the post-shock region (i.e. Comptonized corona) is smaller in
size and hence the contribution of hard photons is less. In the SIMS the disk
accretion rate has an average value $\dot{m}_d = 0.23$ while $\dot{m}_h = 0.09$.
Variabilities are observed in this state (see Figure \ref{fig:Variability}) 
and model fitted parameters during these variabilities are marked with magenta coloured diamonds in Figure
\ref{fig:Accrn} and with white diamonds in the `q'-plot (Figure \ref{fig:Qplot}). 
It is clear from the `q'-plot that during these variabilities the source does not show any significant
change in the flux values.

Then the system reaches the end of the rising phase by entering into the HSS. In the HSS the shock location is a low value of $r_s=14$. This implies that the disk
is more dominant than the halo, with the disk accretion rate $\dot{m}_d = 0.39$ and a low $\dot{m}_h = 0.08$. The corresponding point is marked with a red patch in Figure \ref{fig:Qplot}.

Following the HSS is the decay phase wherein the state change from SIMS to HIMS and finally to LHS.
In the SIMS decay phase (orange patch as shown in Figure \ref{fig:Qplot}) the
accretion rate remains almost constant while the shock location abruptly goes up to
around $r_s=100$. In the HIMS decay phase the shock location gradually starts rising to $r_s=201$. 
Here the halo rate remains constant but the disk rate decreases from $\dot{m}_d = 0.14$ to $\dot{m}_d = 0.02$.
The decay phase of HIMS is marked with transparent yellow coloured patch in
Figure \ref{fig:Qplot} extending from E to F. Finally in the LHS of the decay
phase the shock location rises up-to $r_s=300$ and, $\dot{m}_h = 0.19$ exceeds the disc accretion rate $\dot{m}_d = 0.007$.

\section{Discussion and Conclusions}
\label{dis}

In this paper, we have studied the spectral and temporal variabilities of the black hole source 
IGR J17091$-$3624 during its 2016 outburst. It has been understood very well that based on the variations of 
the thermal and non-thermal emission, the BHs exhibit several spectral states and form a `q'-shaped HID profile. 
Several studies based on theoretical models have looked into these, and explored the accretion 
phenomenon (\citealt{Belloni2005,2010MNRAS.403...61D,Nandi2012,RN2014} and references therein). The two 
component advective flow model \citep{ST95} suggests that the thermal emission is occurring from the 
Keplerian flow. The sub-Keplerian corona inverse Comptonize 
the soft photons resulting in a power-law hard photons distribution. Depending upon the contribution 
from Keplerian and sub-Keplerian flow during the accretion process, BH sources occupy the different spectral 
states. Based on this understanding of the accretion phenomenon, in this paper we have looked into the 
evolution of spectral and temporal characteristics of IGR J17091$-$3624 during its recent outburst in 2016. 

We find that during the rising phase of the outburst, the source occupies hard and hard intermediate states. 
During the hard state, the source exhibits a powerlaw spectrum with the hardness ratio $\sim$0.9 and high fractional rms 
variability in the PDS (as shown in Table \ref{tab:parameters}, Figure \ref{fig:fig3}). The 
broad-band fit using two component flow model implies that the values of shock location and sub-Keplerian halo rate are 
maximum (see Figure \ref{fig:Accrn}). In the HIMS, the contribution of thermal emission increases. This is also evident 
in the decrease of both the hardness ratio 
and fractional rms variability (right panel of Figure \ref{fig:fig3}). The shock location 
and halo rate have decreased while the Keplerian disc accretion 
rate has increased. These factors suggest the rise of thermal emission. Presence of reflection component at higher energies have been 
observed for a few observations in the LHS and HIMS of the rising phase (see also \citealt{Xu17}). In addition to this, we find that the value of cut-off at high energies has a decreasing trend. This is similar to that observed in a few other black hole binaries 
like GX 339$-$4 \citep{Motta2009}. 

An evolution of type C QPO frequencies is also observed from the LHS 
to HIMS. The QPOs observed in {\it XRT} data are weaker. But strong presence of QPOs are shown by \emph{NuSTAR} observations 
(see Figure \ref{fig:Nustar-PDS} and Table \ref{tab:qpo}). Although \citealt{Xu17} has reported about the detection of QPOs in 
{\it NuSTAR} data of this outburst of the source, we understand that they have considered only observations belonging to the rising phase of the outburst. 
Here, in this manuscript we have found additional QPOs for many other observations and also the mHz QPOs which have not been discussed yet. In \citealt{Sree2018} we have also further studied this evolution of QPOs using the propagating oscillation solution \citep{skc2008} of the two component flow model.

Following this, we find that the source enters the SIMS as evident in the decline of disc temperature, hardness ratio and fractional rms variability (see panel h of Figure \ref{fig:fig1} and right panel of Figure \ref{fig:fig3}) w.r.t HIMS. Thus the emission is dominated by that due to the Keplerian 
disc. A higher ratio of 
Keplerian to halo accretion rate is observed (see Figure \ref{fig:Accrn}). 
For a very brief period on day 63 (MJD 57508), the source spectrum becomes softer with 
hardness ratio attaining its minimum value and also a lesser value of fractional rms variability (Figure \ref{fig:fig3}). The spectral softening is also reflected in the increase of Keplerian disc accretion rate to a maximum of 0.39 M$_{Edd}$ which is more than that during the SIMS-rise and decay phases.
We understand that probably this short duration belongs to a HSS and this is unlike to typical
BH sources (see \citealt{Belloni2005,MR06}).
The source later decays through the SIMS, HIMS and LHS, with a reverse trend of change in the Keplerian and halo 
accretion rates, and shock location. The outburst under consideration has a steep rise 
and exponential decay pattern. Hence the states in the decay phase persists for longer duration than the corresponding states 
in the rising phase.

Four broad-band \emph{XRT+NuSTAR} observations (0.5 - 79 keV) have been modelled using two component model during different 
states of the outburst. It provides a better estimation of thermal and non-thermal contributions in the spectra. 
The behaviour of the model parameters are consistent with state transitions as mentioned before. 
Also they are found to be consistent with the values obtained by two component modelling of the {\it XRT} spectra alone 
(see Figure \ref{fig:Accrn}). Thus based on the phenomenological and two component model fits, we understand that 
the source occupies all 
the spectral states in the HID and completes the `q'-profile. 

Previous publications have not been successful in producing a \emph{complete} `q'-diagram for the 2011 outburst of the source. The only published paper which has shown a `q'-diagram is \citealt{Cap12} where the profile was incomplete since the study considered only till the HSS. The Astronomer's Telegram by \citealt{PahariATEL1,PahariATEL2} discussed the source decaying towards its quiescence. But due to poor signal-to-noise ratio of the {\it XRT} data, they could not study the detailed spectral characteristics and hence the `q'-profile during decay phase. In this manuscript, for the 2016 outburst we have been able to understand that the source completes the `q'-diagram based on both phenomenological and two component flow modelling. This has been possible due to the {\it XRT} data having statistically significant count rate throughout the outburst unlike the 2011 outburst.  

An interesting fact is that throughout the entire rising phase of the SIMS of the 2016 outburst, we observe 
variabilities/oscillations in the light-curve (see Figures \ref{fig:LCPDS} and \ref{fig:Variability}) 
until the source entered the decay phase. The time period of oscillations suggest the presence of mHz QPOs. The extracted temporal data 
prove the existence of mHz QPOs for most of the days in the SIMS. Interestingly, all the observations with variability showed 
prominence of \textit{diskbb} in the energy spectra for the phenomenological modelling. Higher 
Keplerian to halo accretion rate is also exhibited as compared to the other states (see bottom panel of Figure \ref{fig:Accrn}). This implies that 
the cause of the variability is essentially thermal in nature. Hence linked to the presence of a Keplerian disc 
that reaches close to the black hole. This is consistent with smaller values of shock location during this state (top panel 
of Figure \ref{fig:Accrn}). The signature of variabilities during day 46.54 of the SIMS has also been reported in 
an Astronomer's Telegram by \citealt{2016ATel.8948....1R}. In 2011 outburst, similar oscillations were observed after the HSS only 
and existed for a long duration \citep{Cap12}. 

Apart from the variabilities during the SIMS, we observe weak signatures of variabilities during the transition from 
HIMS to SIMS, HSS and the initial days of SIMS-decay. There is no clear detection of variability in LHS possibly due to
 less source flux and hence is statistically insignificant. A weak 16 mHz QPO is observed in the \emph{NuSTAR} observation 
during day 31.11 while the source is transiting from HIMS to SIMS. For the observation in HSS, both the \emph{XRT} 
and \emph{NuSTAR} PDS show broad QPO at 21.4 mHz and 20.7 mHz respectively. A very weak peaked component at 0.16 Hz is observed in the latter.

We would like to highlight that when we study outbursting sources as time dependent events, analytical models are not able to address the time evolution of the outbursting events similar to the variability signatures observed in this source. This is due to the fact that all analytical models are applicable in steady state situations only. Using numerical MHD simulations which include radiative cooling processes, might prove fruitful for this. 
In this paper we observed that the various types of heartbeat oscillations observed in this source are associated
with intermediate states. According to two component flow model, the two types of accretion rate (Keplerian and 
sub-Keplerian) becomes comparable during the intermediate states. It might be possible that the radiation coupling between corona and with both types of flow (having different viscous time scale) may play a role to understand fast variabilities observed in this source. Detailed investigation of this
feature using our model is at present beyond the scope of this paper.

The two component spectral modelling shows that the overall mass accretion rate is only 0.39 M$_{Edd}$ during the HSS. Also, in section \ref{hssp}, we did find that the soft flux contribution to the entire spectra is only 33\%. Thus the lack of Keplerian matter might be the reason for the source to occupy a short duration of HSS. This may not be allowing the source to be in HSS for a longer period of time before transitioning into the SIMS of the decay phase. All these factors indicate that possibly the outburst is triggered due to disc instability at the outer edge. And maybe a small amount of 
sub-Keplerian matter can be converted into the Keplerian matter \citep{2010ApJ...710L.147M}. 

We also note that during the beginning of SIMS the spectral data extends only up-to 6 keV. This causes sudden absence 
of high energy flux suggesting that possibly a jet ejection has occurred. Unfortunately radio flares have not 
been observed as the system reaches the SIMS unlike the case in some outbursting sources \citep{FBG04,FHB09,MJ2012,RN2014,RNVS16a}.  

By means of modelling the broad-band observations by \emph{SWIFT} and \emph{NuSTAR} X-ray observatories, we estimate 
the mass of the black hole candidate using the two component flow model. The range of values in which 
the mass varies is found to be 10.62 - 12.33 M$_{\sun}$ including the systematic variation as mentioned earlier. 
This is consistent with the previous estimate of 11.8 - 13.7 M$_{\sun}$ using similar methodology of spectral modelling 
by \citealt{2015ApJ...807..108I}. It has to be clearly noted that in phenomenological models the parameters like T$_{in}$, powerlaw index are the distinct spectral signatures which can be tuned independently along with normalizations. In two-component model the parameters appear in hydrodynamic equations which self-consistently calculate the spectral features. This model has only one normalization and not separate normalizations as in diskbb and powerlaw. The advantage of this model is that mass and the accretion rate of the source self-consistently determine the density and temperature distribution of the flow. This in turn determine the spectral signatures like fraction of inverse-Comptonized black body photons, spectral index etc. So, the model chooses the correct mass of the source to match all the spectral features of all the data sets.

Thus from the phenomenological and two component accretion model fits, we can summarize the following points 
about the 2016 outburst of IGR J17091$-$3624. 
\begin{itemize}
\item The source occupies all the spectral states and completes the `q'-profile in HID. Variation of 
parameters from phenomenological fits and two component model fits corroborate the same. Spectral state evolution based on correlation between the fractional rms variability and hardness ratio, is evident from the Hardness-RMS diagram.
\item The halo accretion rate dominates during the hard and hard-intermediate states while the Keplerian - disc rate dominates the softer states.
\item The size of the Compton corona (shock location) is minimum during the soft state and maximum during the LHS.
\item Presence of reflection component due to ionized material, and decline in cut off energy are observed during the rising phase LHS and HIMS.
\item An evolution of low frequency type C QPOs from 0.15 to 2.15 Hz is observed during the rising phase of LHS and HIMS. 
\emph{NuSTAR} observations show strong signatures of QPOs during the LHS and HIMS.
\item Coherent oscillations/variabilities are exhibited throughout the SIMS and also during the possible HSS. A very weak 
signature is seen during the transition from HIMS to SIMS in the rising phase. QPOs of the order of 20 mHz - 30 mHz are 
found during the SIMS. A 20.7 mHz broad QPO and a weak peaked component at 0.16 Hz exist during the possible HSS.
\item Even in the presence of variabilities, the source completes the `q'-profile in its HID.
\item Mass of the source is estimated to be in the range of 10.62 - 12.33 M$_{\sun}$.
\end{itemize}

\acknowledgments
We are thankful to the reviewer for his valuable comments and suggestions which have helped in improving the manuscript.

This research has made use of the data obtained through High Energy Astrophysics Science 
Archive Research Center on-line service, provided by NASA/GSFC. \\RD is grateful to the 
support provided by the Vice Chancellor and Dean-SOE of DSU, and AN thanks GD, SAG; DD, PDMSA 
and Director, ISAC for encouragement and continuous support to carry out this research.


\begin{thebibliography}{}
\bibitem[Altamirano \& Belloni (2012)]{AB2012} {Altamirano} D.,  {Belloni} T.,  2012, 747, L4

\bibitem[Altamirano et al. (2011)]{Alt11} {Altamirano} D., Belloni, T., Linares, M., et al.\ 2011, \apjl, 742, L17


\bibitem[Arnaud (1996)]{Arn96} {Arnaud} K.~A.,  1996, in {Jacoby} G.~H.,  {Barnes} J.,  eds,  Astronomical Society of the Pacific Conference Series Vol. 101, Astronomical Data Analysis Software and Systems V. p.~17

\bibitem[Belloni \& Hasinger (1990)]{BH1990} Belloni T.~M.,  Hasinger G.,  1990, \aap, 230, 103

\bibitem[Belloni et al. (2001)]{2001A&A...372..551B} {Belloni} T.,  {M{\'e}ndez} M.,   {S{\'a}nchez-Fern{\'a}ndez} C.,  2001, 372, 551

\bibitem[Belloni et~al. (2005)]{Belloni2005} Belloni T.,  Homan J.,  Casella P.,   et~al., 2005, \aap, 440, 207

\bibitem[Belloni et~al. (2011)]{2011BASI...39..409B} {Belloni} T.~M.,  {Motta} S.~E.,   {Mu{\~n}oz-Darias} T.,  2011, Bulletin of
  the Astronomical Society of India, 39, 409

\bibitem[Brocksopp et~al. (2002)]{Brock2002} Brocksopp C.,  Fender R.~P.,  McCollough M.,   et~al., 2002, \mnras, 331, 765

\bibitem[Burrows et~al. (2005)]{2005SSRv..120..165B} {Burrows} D.~N.,  et~al., 2005, {120, 165}

\bibitem[Capitanio et~al. (2012)]{Cap12} {Capitanio} F.,  {Del Santo} M.,  {Bozzo} E.,  {Ferrigno} C.,  {De Cesare} G., {Paizis} A.,  2012, {422, 3130}

\bibitem[Capitanio et al. (2013)]{Cap13} {Capitanio} F.,  {Del Santo} M.,  {Bozzo} E.,  {Ferrigno} C.,  {De Cesare} G., {Paizis} A.,  2013, preprint, arXiv 1302.3485

\bibitem[Casella et~al. (2004)]{Casella2004} Casella P.,  Belloni T.,  Homan J.,   Stella L.,  2004, \aap, 426, 587

\bibitem[Chakrabarti (1996)]{1996ApJ...464..664C} Chakrabarti S.~K.,  1996, 464, 664

\bibitem[Chakrabarti \& Das (2004)]{2004MNRAS.349..649C} {Chakrabarti} S.~K.,  {Das} S.,  2004, {349, 649}

\bibitem[Chakrabarti \& Titarchuk (1995)]{ST95} Chakrabarti S.,  Titarchuk L.,  1995, \apj, 455, 623

\bibitem[Chakrabarti et al. (2008)]{skc2008} Chakrabarti, S.~K., Debnath, D., Nandi, A., \& Pal, P.~S.\ 2008, \aap, 489, L41 

\bibitem[Chattopadhyay \& Chakrabarti (2011)]{2011IJMPD..20.1597C} {Chattopadhyay} I.,  {Chakrabarti} S.~K.,  2011, {20, 1597}

\bibitem[Chen et~al. (1997)]{Chen97} {Chen} W.,  {Shrader} C.~R.,   {Livio} M.,  1997, {491, 312}

\bibitem[Corral-Santana et~al. (2016)]{2016A&A...587A..61C} {Corral-Santana} J.~M.,  {Casares} J.,  {Mu{\~n}oz-Darias} T.,  {Bauer} F.~E.,
  {Mart{\'{\i}}nez-Pais} I.~G.,   {Russell} D.~M.,  2016, {587, A61} 

\bibitem[Court et~al. (2016)]{2016ATel.8858....1C}{Court} J.~M.~C.,  {Motta} S.~E.,   {Altamirano} D.,  2016, The Astronomer's Telegram, {8858}

\bibitem[Court et~al. (2017)]{2017arXiv170309572C} {Court} J.~M.~C.,  {Altamirano} D.,  {Pereyra} M.,  {Boon} C.~M.,  {Yamaoka}
  K.,  {Belloni} T.,  {Wijnands} R., {Pahari} M.,  2017, MNRAS, 468, 4748

\bibitem[Das (2007)]{2007MNRAS.376.1659D} {Das} S.,  2007, {376, 1659}

\bibitem[Dunn et~al. (2010)]{2010MNRAS.403...61D}{Dunn} R.~J.~H.,  {Fender} R.~P.,  {K{\"o}rding} E.~G.,  {Belloni} T., {Cabanac} C.,  2010, {403, 61}

\bibitem[Egron et~al. (2016)]{2016ATel.8821....1E} {Egron} E.,  et~al., 2016, The Astronomer's Telegram, {8821}

\bibitem[Fender et~al. (2004)]{FBG04} Fender R.~P.,  Belloni T.,   Gallo E.,  2004, \mnras, 355, 1105

\bibitem[Fender et~al. (2009)]{FHB09} Fender R.~P.,  Homan J.,   Belloni T.,  2009, \mnras, 396, 1307

\bibitem[Greiner et~al. (2016)]{2016ATel.8795....1G} {Greiner} J.,  {Bolmer} J.,  {Gandhi} P.,  {Altamirano} D.,  {Charles} P.~A., {Court} J.~M.,  {Kann} D.~A.,   {Walton} D.~J.,  2016, The Astronomer's Telegram, {8795}

\bibitem[Grinberg et~al. (2016)]{2016ATel.8761....1G} {Grinberg} V.,  et~al., 2016, The Astronomer's Telegram, {8761}

\bibitem[Harrison et~al. (2013)]{harrison2013nuclear} {Harrison} F.~A.,  et~al., 2013, {770, 103}

\bibitem[Homan \& Belloni (2005)]{HB2005} Homan J.,  Belloni T.,  2005, \apss, 300, 107

\bibitem[Homan et~al. (2001)]{2001ApJS..132..377H} {Homan} J.,  {Wijnands} R.,  {van der Klis} M.,  {Belloni} T.,  {van Paradijs}  J.,  {Klein-Wolt} M.,  {Fender} R.,   {M{\'e}ndez} M.,  2001, {132, 377}

\bibitem[Iyer \& Nandi (2013)]{Iyer13} {Iyer} N.,  {Nandi} A.,  2013, in {Das} S.,  {Nandi} A.,   {Chattopadhyay} I.,
  eds,  Astronomical Society of India Conference Series Vol. 8, Astronomical
  Society of India Conference Series. pp 79--83 

\bibitem[Iyer et~al. (2015)]{2015ApJ...807..108I} {Iyer} N.,  {Nandi} A.,   {Mandal} S.,  2015, {807, 108}

\bibitem[King et~al. (2012)]{king2012extreme} {King} A.~L.,  et~al., 2012, {746, L20}

\bibitem[Koljonen et al. (2013)]{Koljo13} Koljonen, K.~I.~I., McCollough, M.~L., Hannikainen, D.~C., \& Droulans, R.\ 2013, \mnras, 429, 1173 

\bibitem[Krimm et~al. (2011)]{2011ATel.3144....1K} {Krimm} H.~A.,  et~al., 2011, The Astronomer's Telegram, {3144}

\bibitem[Kuulkers et~al. (2003)]{Kuul03} {Kuulkers} E.,  {Lutovinov} A.,  {Parmar} A.,  {Capitanio} F.,  {Mowlavi} N., {Hermsen} W.,  2003, The Astronomer's Telegram, {149}

\bibitem[Kuznetsov et~al. (1997)]{Kuz97} {Kuznetsov} S.,  et~al., 1997, {292, 651}

\bibitem[Makishima et~al. (1986)]{Maki86} Makishima K.,  Maejima Y.,  Mitsuda K.,   et~al., 1986, \apj, 308, 635

\bibitem[Mandal \& Chakrabarti (2005)]{2005A&A...434..839M} {Mandal} S.,  {Chakrabarti} S.~K.,  2005, {434, 839}

\bibitem[Mandal \& Chakrabarti (2010)]{2010ApJ...710L.147M} {Mandal} S.,  {Chakrabarti} S.~K.,  2010, {710, L147}

\bibitem[McClintock \& Remillard (2006)]{MR06} McClintock J.~E.,  Remillard R.~A.,  2006, `Black hole binaries', Compact Stellar X-ray sources, edited by Lewin W. H. G. and M. van der Klis

\bibitem[Miller-Jones et~al. (2012)]{MJ2012} {Miller-Jones} J.~C.~A.,  et~al., 2012, {421, 468}

\bibitem[Miller et~al. (2016)]{2016ATel.8742....1M} {Miller} J.~M.,  {Reynolds} M.,  {Kennea} J.,  {King} A.~L.,   {Tomsick} J., 2016, The Astronomer's Telegram, {8742}

\bibitem[Mitsuda et~al. (1984)]{1984PASJ...36..741M} {Mitsuda} K.,  et~al., 1984, \pasj, {36, 741}

\bibitem[Motta et~al. (2009)]{Motta2009} {Motta} S.,  {Belloni} T.,   {Homan} J.,  2009, {400, 1603}

\bibitem[Motta et~al. (2012)]{Motta2012} Motta S.,  Homan J.,  Munoz-Darias T.,   et~al., 2012, \mnras, 427, 595

\bibitem[Muno et~al. (1999)]{1999ApJ...527..321M} {Muno} M.~P.,  {Morgan} E.~H.,   {Remillard} R.~A.,  1999, {527, 321}

\bibitem[Nandi et~al. (2012)]{Nandi2012} Nandi A.,  Debnath D.,  Mandal S.,   Chakrabarti S.~K.,  2012, \aap, 542, 56

\bibitem[Pahari et~al. (2012a)]{PahariATEL1} {Pahari} M.,  {Bhattacharyya} S.,   {Yadav} J.~S.,  2012a, The Astronomer's Telegram, {4282}

\bibitem[Pahari et~al. (2012b)]{PahariATEL2} {Pahari} M.,  {Bhattacharyya} S.,   {Yadav} J.~S.,  2012b, The Astronomer's Telegram, {4283}

\bibitem[Pahari et~al. (2014)]{Pah2014} {Pahari} M.,  {Yadav} J.~S.,   {Bhattacharyya} S.,  2014, {783, 141}

\bibitem[Pahari et al. (2017)]{Pahari17} Pahari, M., Antia, H.~M., Yadav, J.~S., et al.\ 2017, \apj, 849, 16 

\bibitem[Radhika \& Nandi (2014)]{RN2014} {Radhika} D.,  {Nandi} A.,  2014, {54, 1678}

\bibitem[Radhika et~al. (2016a)]{RNVS16a} {Radhika} D.,  {Nandi} A.,  {Agrawal} V.~K.,   {Seetha} S.,  2016a, {460, 4403}

\bibitem[Radhika et~al. (2016b)]{RNVS16b} {Radhika} D.,  {Nandi} A.,  {Agrawal} V.~K.,   {Mandal} S.,  2016b, {462, 1834}

\bibitem[Rao \& Vadawale (2012)]{RV2012} {Rao} A.,  {Vadawale} S.~V.,  2012, {757, L12}

\bibitem[Rebusco et~al. (2012)]{Rebusco12} {Rebusco} P.,  {Moskalik} P.,  {Klu{\'z}niak} W.,   {Abramowicz} M.~A.,  2012, {540, L4}

\bibitem[Remillard \& McClintock (2006)]{2006ARA&A..44...49R} {Remillard} R.~A.,  {McClintock} J.~E.,  2006, {44, 49}

\bibitem[Reynolds et~al. (2016)]{2016ATel.8948....1R} {Reynolds} M.,  {Miller} J.,   {King} A.,  2016, The Astronomer's Telegram, {8948}

\bibitem[Rodriguez et~al. (2011)]{rodriguez2011first} {Rodriguez} J.,  {Corbel} S.,  {Caballero} I.,  {Tomsick} J.~A.,  {Tzioumis}  T.,  {Paizis} A.,  {Cadolle Bel} M.,   {Kuulkers} E.,  2011, {533, L4}

\bibitem[Ross \& Fabian (1993)]{RF93} Ross, R.~R., \& Fabian, A.~C.\ 1993, \mnras, 261, 74 

\bibitem[Savolainen (2012)]{Sav12} Savolainen, P.\ 2012, X-ray Binaries.~Celebrating 50 Years Since the Discovery of Sco X-1, 52 


\bibitem[Shakura \& Sunyaev (1973)]{1973A&A....24..337S} {Shakura} N.~I.,  {Sunyaev} R.~A.,  1973, \aap, {24, 337}

\bibitem[Sobolev (1975)]{1975lpsa.book.....S} {Sobolev} V.~V.,  1975, {Light scattering in planetary atmospheres}

\bibitem[Sreehari et al.(2018)]{Sree2018} Sreehari, H., Nandi, A., Radhika, D., Iyer, N., \& Mandal, S.\ 2018, Journal of Astrophysics and Astronomy, 39, \#5 

\bibitem[Tanaka \& Lewin (1995)]{1995xrbi.nasa..126T} {Tanaka} Y.,  {Lewin} W. H.~G.,  1995, X-ray Binaries, pp 126--174

\bibitem[Tanaka \& Shibazaki (1996)]{1996ARA&A..34..607T} {Tanaka} Y.,  {Shibazaki} N.,  1996, {34, 607}

\bibitem[Tetarenko et~al. (2016)]{2016ApJS..222...15T} {Tetarenko} B.~E.,  {Sivakoff} G.~R.,  {Heinke} C.~O.,   {Gladstone} J.~C., 2016, {222, 15}

\bibitem[Tsunemi et~al. (1989)]{1989ApJ...337L..81T} {Tsunemi} H.,  {Kitamoto} S.,  {Okamura} S.,   {Roussel-Dupre} D.,  1989, {337, L81}

\bibitem[Wilms et~al. (2000)]{2000ApJ...542..914W} {Wilms} J.,  {Allen} A.,   {McCray} R.,  2000, {542, 914}

\bibitem[Xu et~al. (2017)]{Xu17} {Xu} Y.,  et~al., 2017, ApJ, 851, 103

\bibitem[in't Zand et~al. (2003)]{2003ATel..160....1I} {in't Zand} J.~J.~M.,  {Heise} J.,  {Lowes} P.,   {Ubertini} P.,  2003, The
  Astronomer's Telegram, {160}


\bibitem[Zhang et al.(2014)]{Zhang2014} Zhang, Z., Qu, J.~L., Gao, H.~Q., et al.\ 2014, \aap, 569, A33 

\end{thebibliography}

\appendix
\section{Appendix}

\begin{longtable}{cccccc}
\caption{Log of \emph{SWIFT-XRT} and \emph{NuSTAR} observations considered for analysis}\\
\hline\\
\multicolumn{2}{c}{Observation ID} & Date & Time & MJD & Exposure time \\
\cline{1-2}\\
\textit{SWIFT}&\textit{\emph{NuSTAR}}&&&&(in sec)\\
\hline\\
00031921092&	&2016-02-27& 08:24:58&	57445.35 &999.478\\  
00031921093&	&2016-02-29& 06:31:49&	57447.27 &994.458\\  
00031921095&	&2016-03-03& 22:40:19&	57450.94 &399.444\\
00031921097&	&2016-03-05& 12:49:43&	57452.53 &849.457\\
&80001041002	&2016-03-07& 02:01:08&	57454.08 &43293\\
00031921101&	&2016-03-09& 06:17:18&	57456.26 &1958.855\\ \hline\\
00031921104&    &2016-03-12& 13:56:47& 57459.58 &1954.263\\
&80202014002    &2016-03-12& 14:11:08& 57459.59 &20238\\ \hline\\
00031921105&	&2016-03-13& 17:14:59&	57460.71 &934.618 \\ \hline\\
&80202014004    &2016-03-14& 19:26:08& 57461.80 &20698\\
00031921106&    &2016-03-14 &21:43:26& 57461.90 &1049.610\\ \hline\\
00031921110&	&2016-03-20& 21:09:58&	57467.88 &1004.331\\
00031921111&	&2016-03-21& 00:39:40&	57468.02 &974.618\\
00031921112&	&2016-03-22& 19:36:45&	57469.81 &1089.453\\
00031921113&	&2016-03-23& 19:33:58&	57470.81 &849.447\\
00031921117&	&2016-03-27& 14:25:14&	57474.60 &669.621\\
&80202014006	&2016-03-29& 02:41:08&	57476.11 &48393\\
00031921121&	&2016-04-01& 20:11:58&	57479.84 &1044.448\\
00031921124&	&2016-04-04& 16:46:58&	57482.69 &979.446\\
00031921132&	&2016-04-12& 17:45:58&	57490.74 &904.443\\
00031921133&	&2016-04-13& 13:05:46&	57491.54 &849.446\\
00031921134&	&2016-04-14& 22:50:42&	57492.95 &849.617\\
00031921142&	&2016-04-23& 15:27:58&	57501.64 &981.146\\ 
00031921143&   &2016-04-24& 09:07:57&  57502.38 &964.443\\
00031921144&   &2016-04-27& 18:34:58&  57505.77 &869.621\\
00031921145&   &2016-04-28& 13:54:58&  57506.57 &979.313\\ \hline\\
&80202015002   &2016-04-30& 11:26:08& 57508.47  &39113\\
00081917001&   &2016-04-30& 21:26:45&  57508.89 &1859.223\\  \hline\\
00031921147&	&2016-05-05& 11:26:39&	57513.47 &789.447\\
00031921150&	&2016-05-08& 19:30:58&	57516.81 &774.617\\
00031921153&	&2016-05-11& 19:00:58&	57519.79 &974.617\\
00034543003&	&2016-05-13& 07:35:58&	57521.31 &954.608\\
00034543006&	&2016-05-16& 12:09:58&	57524.50 &959.607\\
00031921156&	&2016-05-16& 18:31:57&	57524.77 &909.621\\
00034543009&	&2016-05-19& 13:33:58&	57527.56 &984.609\\
00034543013&	&2016-05-24& 21:17:58&	57532.88 &974.556\\
00031921160&	&2016-05-25& 00:47:58&	57533.03 &504.441\\ \hline\\
&80202015004    &2016-05-26& 15:51:08& 57534.66 &36340\\
00081917002&    &2016-05-26& 16:11:52& 57534.67 &1748.128\\ \hline \\
00031921161&	&2016-05-30& 08:28:45&	57538.35 &249.606\\
00031921162&	&2016-05-31& 01:51:29&	57539.07 &864.657\\
00031921163&	&2016-06-01& 01:42:58&	57540.07 &924.617\\
00031921166&	&2016-06-04& 01:26:57&	57543.06 &989.449\\
00031921170&   &2016-06-21& 23:58:58&  57560.99 &1252.835\\
00031921173&   &2016-06-30& 17:04:58&  57569.71 &364.461\\
00031921174&   &2016-07-02& 20:01:58&  57571.83 &994.620\\
00031921175&   &2016-07-04& 11:39:57&  57573.48 &1209.619\\
00031921176&   &2016-07-06& 16:19:58&  57575.68 &919.784\\
00031921178&   &2016-07-11& 17:32:58&  57580.73 &444.443\\
00031921179&   &2016-07-12& 00:23:57&  57581.01 &1173.715\\
00031921180&   &2016-07-14& 12:43:59&  57583.53 &1319.617\\
00031921183&   &2016-07-22& 15:22:58&  57591.64 &1428.042\\
00031921184&   &2016-07-24& 22:48:58&  57593.95 &969.272\\
00031921185&   &2016-07-26& 11:39:57&  57595.48 &1439.619\\
00031921187&   &2016-08-12& 00:51:57&  57612.03 &1519.016\\
00031921189&   &2016-08-17& 19:41:58&  57617.82 &1559.234\\
\hline
\label{tab:obs}
\end{longtable}

\begin{table*}
	\centering
	\caption{Broadband fit models using various phenomenological models and corresponding $\chi^2$ values}
	\label{tab:bbfit}
	\begin{tabular}{lcccr} 
		\hline
		MJD &Observatory& State&Model & $\chi ^2/DOF$\\
		\hline
		57454&{\it NuSTAR}&LHS &phabs(powerlaw)& 1221.16/990=1.23\\
		     && &phabs(cutoffpl) &1198.05/989=1.21\\
		     && &phabs(compTT) &1088.71/988=1.10\\
		     && &phabs(ireflect*cutoffpl)&1011.10/987=1.02\\
		     && &phabs(two-component)&1183.82/987=1.19\\
		     && &phabs(smedge*two-component)&1131.39/985=1.14\\
		     && &phabs(two-component*highecut)&1139.47/986=1.15\\
		\hline
		57459&{\it NuSTAR+XRT}&LHS$->$HIMS &phabs(powerlaw) &1405.42/1221=1.15\\
		     && &phabs(cutoffpl) &1289.63/1220=1.05\\
		     && &phabs(compTT) &1306.39/1219=1.07\\
		     && &phabs(ireflect*cutoffpl)&1234.81/1218=1.01\\
		     && &phabs(two-component)&1240.55/1087=1.14\\
		     && &phabs(smedge*two-component)&1243.52/1086=1.14\\
		     && &phabs(two-component*highecut)&1202.27/1086=1.08\\		     
		\hline
		57461&{\it NuSTAR+XRT}&HIMS &phabs(powerlaw) &1475.41/1116=1.32\\
		     && &phabs(cutoffpl) &1290.67/1115=1.15\\
		     && &phabs(compTT) &1328.03/1114=1.19\\
		     && &phabs(ireflect*cutoffpl)&1202.27/1113=1.08\\
		     && &phabs(two-component)&1567.62/1103=1.42\\
		     && &phabs(smedge*two-component)&1534.64/1100=1.39\\
     	     && &phabs(two-component*highecut)&1227.02/1099=1.11\\			     
		\hline
		57476&{\it NuSTAR}&SIMS &phabs(diskbb+powerlaw) &1064.79/814=1.30\\
		     && &phabs(diskbb+cutoffpl)&1084.38/813=1.33\\
		     && &phabs(diskbb+compTT) &1047.77/812=1.29\\
		     && &phabs(diskbb+ireflect*cutoffpl)&909.25/811=1.12\\
		     && &phabs(two-component)&1667.30/814=2.04\\
		     && &phabs(smedge*two-component)&1037.44/810=1.28\\
		     && &phabs(smedge*smedge*two-component)&894.88/808=1.10\\
		\hline
		57508&{\it NuSTAR+XRT}&HSS &phabs(diskbb+powerlaw) &987.61/969=1.01\\
		     && &phabs(diskbb+cutoffpl) &989.07/969=1.02\\
		     && &phabs(diskbb+compTT) &996.02/967=1.03\\
		     && &phabs(diskbb+ireflect*cutoffpl) &967.07/966=1.00\\
		     && &phabs(smedge*two-component)&1558.65/967=1.61\\
		     && &phabs(smedge*smedge*two-component)&1036.09/964=1.07\\
		\hline
		57534&{\it NuSTAR+XRT}&SIMS-decay &phabs(diskbb+powerlaw) &1000.36/946=1.06\\
		     && &phabs(diskbb+cutoffpl) &1002.35/945=1.06\\
		     && &phabs(diskbb+compTT)&1007.78/944=1.06\\
		     && &phabs(diskbb+ireflect*cutoffpl)&970.46/943=1.02\\
		     && &phabs(two-component)&1201.26/859=1.39\\
		     && &phabs(smedge*two-component)&949.16/854=1.11\\
        \hline 
		
     \end{tabular}
\end{table*}

\nocite{*}


\end{document}